\documentclass[twocolumn]{emulateapj}
\usepackage{graphicx}
\usepackage{amssymb}
\usepackage{amsmath}
\usepackage{subfigure}
\usepackage{multirow}
\usepackage{threeparttable}
\usepackage{array}
\usepackage{natbib}
\usepackage{xcolor}
\usepackage{subfigure}
\usepackage{hyperref}

\hypersetup{
    colorlinks=true,
    linkcolor=red,
    filecolor=magenta,      
    urlcolor=cyan,
}

\def\MgX{{\rm Mg} {\sc x}}

\def\NeVIII{{\rm Ne} {\sc viii}}

\def\OVI{{\rm O} {\sc vi}}
\def\OVII{{\rm O} {\sc vii}}
\def\OVIII{{\rm O} {\sc viii}}

\def\HI{{\rm H} {\sc i}}

\def\kms{~\rm km~s^{-1}}      
\def\cmsq{~\rm cm^{-2}}

\newcommand{\red}{\textcolor{black}}

\begin{document}
\title{The Mass and Absorption Columns of Galactic Gaseous Halos II -- The High Ionization State Ions}
\author{Zhijie Qu and Joel N. Bregman}
\affil{Department of Astronomy, University of Michigan, Ann Arbor, MI 48104, USA}
\email{quzhijie@umich.edu}
\email{jbregman@umich.edu}
\begin{abstract}
The high ionization-state ions trace the hot gases in the universe, of which gaseous halos around galaxies are a major contributor.
Following \citet{Qu:2018aa}, we calculate the gaseous halo contribution to the observed column density distributions for these ions by convolving the gaseous halo model with the observed stellar mass function.
The predicted column density distribution reproduces the general shape of the observed column density distribution -- a broken power law with the break point at $\log N=14.0$ for {\OVI}.
Our modeling suggests that the high column density systems originate from galaxies for which the virial temperature matches the temperature of the ionization fraction peak. 
Specifically, this mass range is $\log M_\star=8.5-10$ for {\OVI}, $\log M_\star=9.5-10.5$ for {\NeVIII}, and higher for higher ionization state ions (assuming $T_{\rm max}=2T_{\rm vir}$).
A comparison with the observed {\OVI} column density distribution prefers a large radius model, where the maximum radius is twice the virial radius.
This model may be in conflict with the more poorly defined {\NeVIII} column density distribution, suggesting further observations are warranted.
The redshift evolution of the high column density systems is dominated by the change of the cosmic star formation rate, which decreases from $z=1.0$ to the local universe.
Some differences at lower columns between our models and observations indicates that absorption by the intra-group (cluster) medium and intergalactic medium are also contributors to the total column density distributions. 
\end{abstract}

\maketitle

\section{Introduction}
During the formation and evolution of galaxies, gases in the cool intergalactic medium (IGM; $T \sim 10^4\rm~K$) fall into dark matter halos and are heated to form the warm-hot intergalactic medium (WHIM; $10^5 - 10^7\rm~ K$) by accretion shocks and various galactic feedback processes \citep{Weinberg:1997aa, Cen:1999aa}.
These warm-hot gases could account for $30-40\%$ of the total baryon contents and exist in different forms: the hot galactic gaseous halo; the intra-group (cluster) medium; and the cosmic web \citep{Cen:2006aa}.
However, these warm-hot gases are difficult to detect because of their low densities and high temperatures. 


Currently, direct X-ray imaging can detect the intra-group (cluster) medium extending to about the virial radius, while for isolated galaxies, X-ray imaging can only detect the emission from hot gases surrounding nearby massive galaxies within about $50\rm~kpc$ \citep{Anderson:2010aa, Bogdan:2013aa, Goulding:2016aa}.
An alternative detection approach is to stack microwave images to measure the average Sunyaev-Zel'dovich (SZ) effect, which is useful for systems with masses above $\log M_{\rm h}=12.3$ \citep{Lim:2018aa}.
Stacking galaxy pairs also reveals the existence of hot cosmic filaments between galaxy pairs within $\approx 15~\rm Mpc$, which account for $20- 30 \%$ of the total baryonic content of the universe \citep{de-Graaff:2017aa, Tanimura:2017aa}.
However, smaller galaxies ($<L^*$) have too weak an SZ signal, even in a stack ($<5\sigma$), to measure a useful constraint on the mass or temperature.
A third way to detect the hot gaseous component in the universe is by measuring absorption lines from high ionization state ions towards background AGNs -- {\OVI}, {\NeVIII}, {\OVII}, {\MgX}, and {\OVIII} -- which can trace the gas with temperatures from $10^5\rm~K$ ({\OVI}) to $5\times 10^6\rm~ K$ ({\OVIII}).

In the past decade, several observational studies of these ions have constrained the column density distributions and the cosmic abundances \citep{Meiring:2013aa, Savage:2014aa, Danforth:2016aa, Frank:2018aa}.
The lithium-like ions {\OVI}, {\NeVIII} and {\MgX} have strong resonant doublets in the far-ultraviolet (FUV) band with wavelengths of $1031.9\rm~\AA$, $770.4\rm~\AA$, and $609.8\rm~\AA$, respectively (for the strongest of the doublet lines). 
Using the current FUV instruments (i.e., {\it Hubble Space Telescope}/Cosmic Origin Spectrograph; {\it HST}/COS), these ions are detectable in the redshift range of around $0.1 - 0.7$, $0.5-1.3$, and $0.9-1.9$. 
{\OVI} has the largest sample among these high ionization state ions because of the low redshift and the high abundance of oxygen \citep{Tripp:2008aa, Thom:2008aa}. 
{\NeVIII} and {\MgX} are more difficult to detect because they are less abundant than oxygen and require a higher continuum S/N ratio ($>20$). 
Currently, fewer than 10 sightlines have reported {\NeVIII} absorption lines (see the summary in \citealt{Pachat:2017aa} and references therein), while there is only one intervening {\MgX} detection \citep[in the sightline toward LBQS 1435-0134;][]{Qu:2016aa}. 
{\OVII} and {\OVIII} occur in the X-ray band and have resonant lines at $21.60\rm~\AA$ and $18.96\rm~\AA$. 
The detection of {\OVII} and {\OVIII} absorption lines is limited to the Milky Way (MW; \citealt{Nevalainen:2017aa}), as detections around external galaxies are still controversial \citep{Fang:2010aa, Nicastro:2016aa}. 
A larger sample of {\OVII} and {\OVIII} will only be feasible with next-generation X-ray telescopes (e.g., {\it Arcus}; \citealt{Smith:2016aa}).

Absorption lines from high ionization state ions make it possible to study the connection between galaxies and their halos. 
For example, {\OVI} is more frequently around star forming galaxies, using the COS-Halos sample \citep{Tumlinson:2011aa, Werk:2016aa}.
This phenomenon is also confirmed for star forming dwarf galaxies, which host {\OVI} absorption of $\approx 10^{14} \cmsq$ \citep{Johnson:2017aa}. 

However, the contributions of gaseous halo to the cosmic high ionization state ions remain as a question, since it is controversial regarding the total baryonic content and the dominant component of gaseous halos. 
Both the cool gas ($\approx 10^4-10^5\rm~K$) and the hot gas ($\approx 10^6\rm~K$) are claimed to account for the missing bayons in $L^*$ galaxies \citep{Gupta:2012aa, Werk:2014aa, Nicastro:2016aa, Prochaska:2017aa}. 
Meanwhile, there are other observational studies and theoretical predictions that the total amount of the gases in halos is $\lesssim 30\%$ of the total baryon mass for $L^*$ galaxies \citep{Miller:2015aa, Schaller:2015aa, Li:2017aa, Bregman:2018aa}.

In \citet[][hereafter, QB18]{Qu:2018aa}, we proposed a semi-analytic galactic gaseous halo model (GGHM) that is consistent with most observations of the high ionization state ions.
This model enables us to consider the contribution to the cosmic high ionization state ions due to galactic gaseous halos.
In this paper, we estimate the column density distribution originating from the galactic gaseous halo by combining the GGHM model with the spatial density of galaxies (i.e., the stellar mass function; SMF).
By comparing the predicted column density distribution to observations, we obtain the relative contribution due to the galaxy and constrain the GGHM model.
In Section 2, we give a brief summary of the GGHM model and the assumptions required to predict the column density distribution. 
In Section 3, we find the preferred model parameters, while in Section 4, we discuss our results and implications.

\begin{figure*}
\begin{center}
\subfigure{
\includegraphics[width=0.33\textwidth]{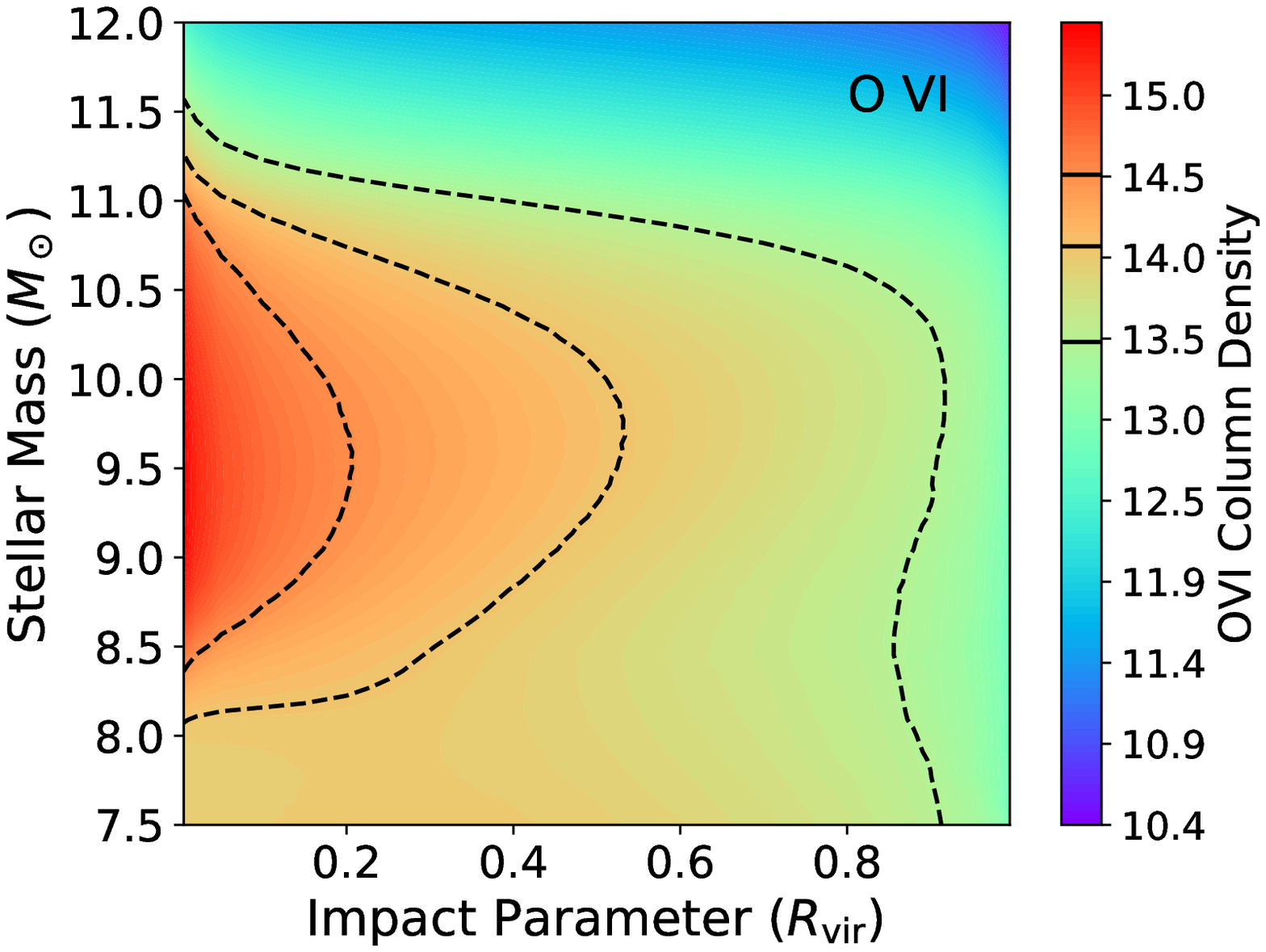}
\includegraphics[width=0.33\textwidth]{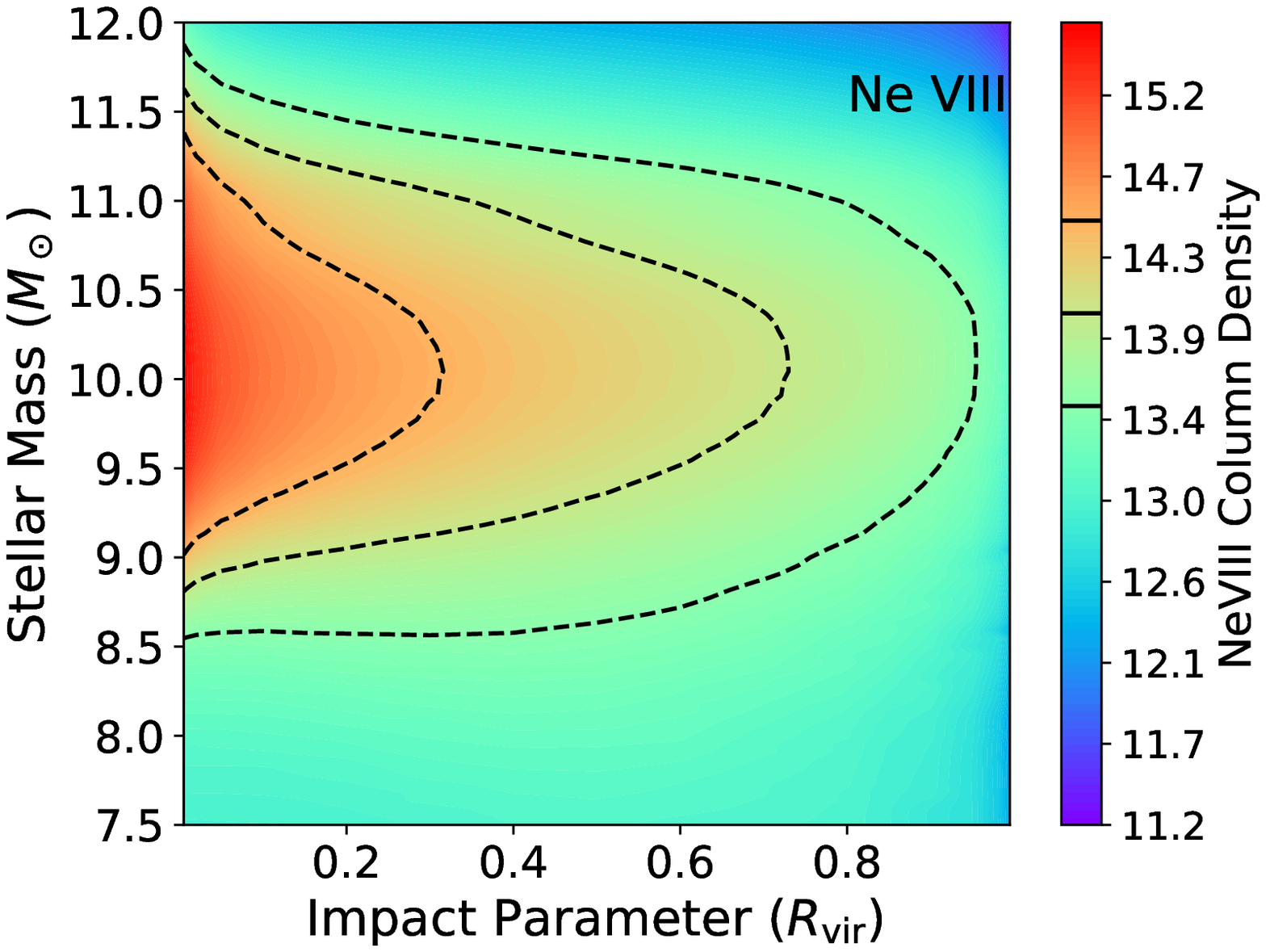}
\includegraphics[width=0.33\textwidth]{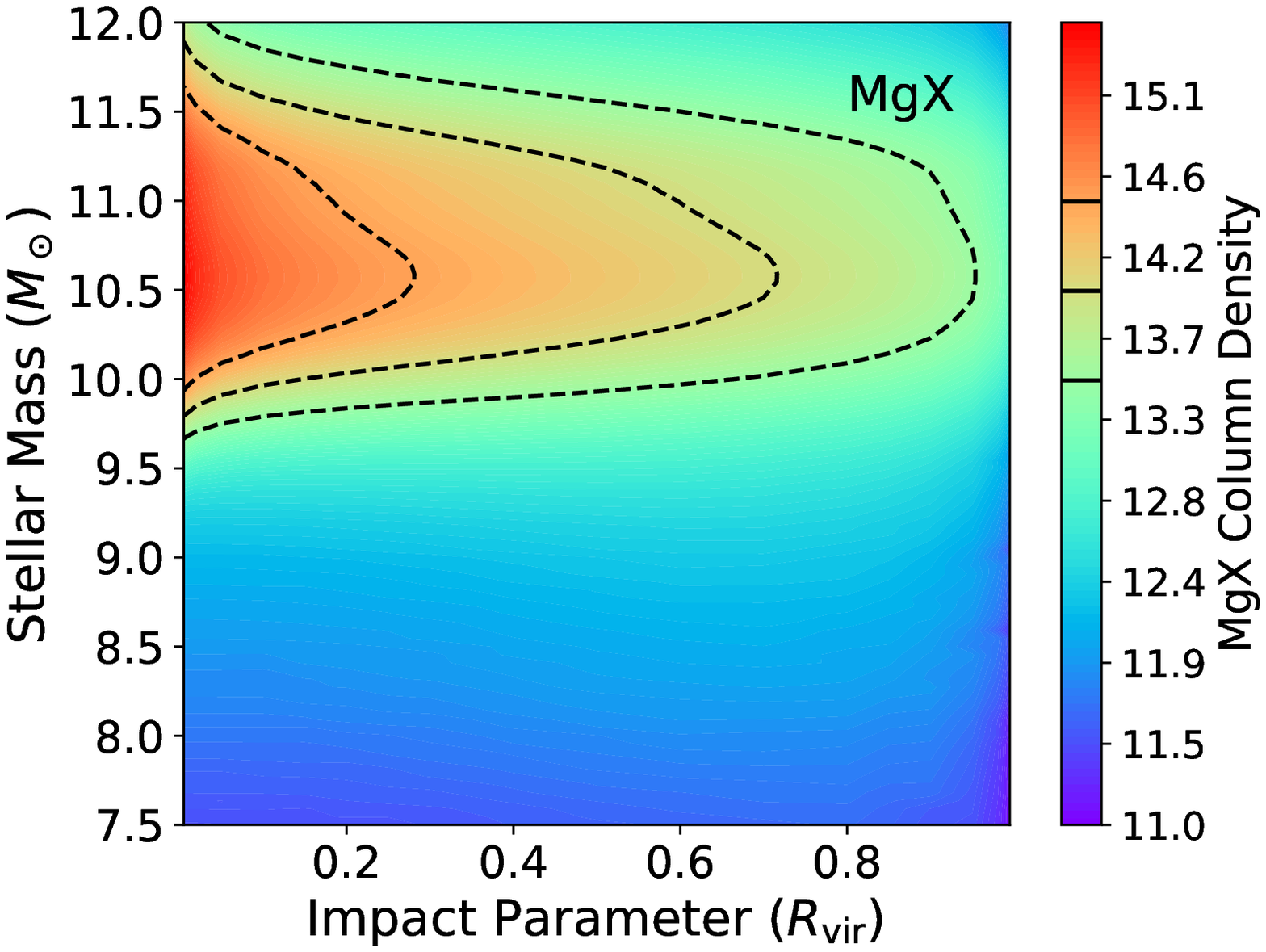}
}
\subfigure{
\includegraphics[width=0.48\textwidth]{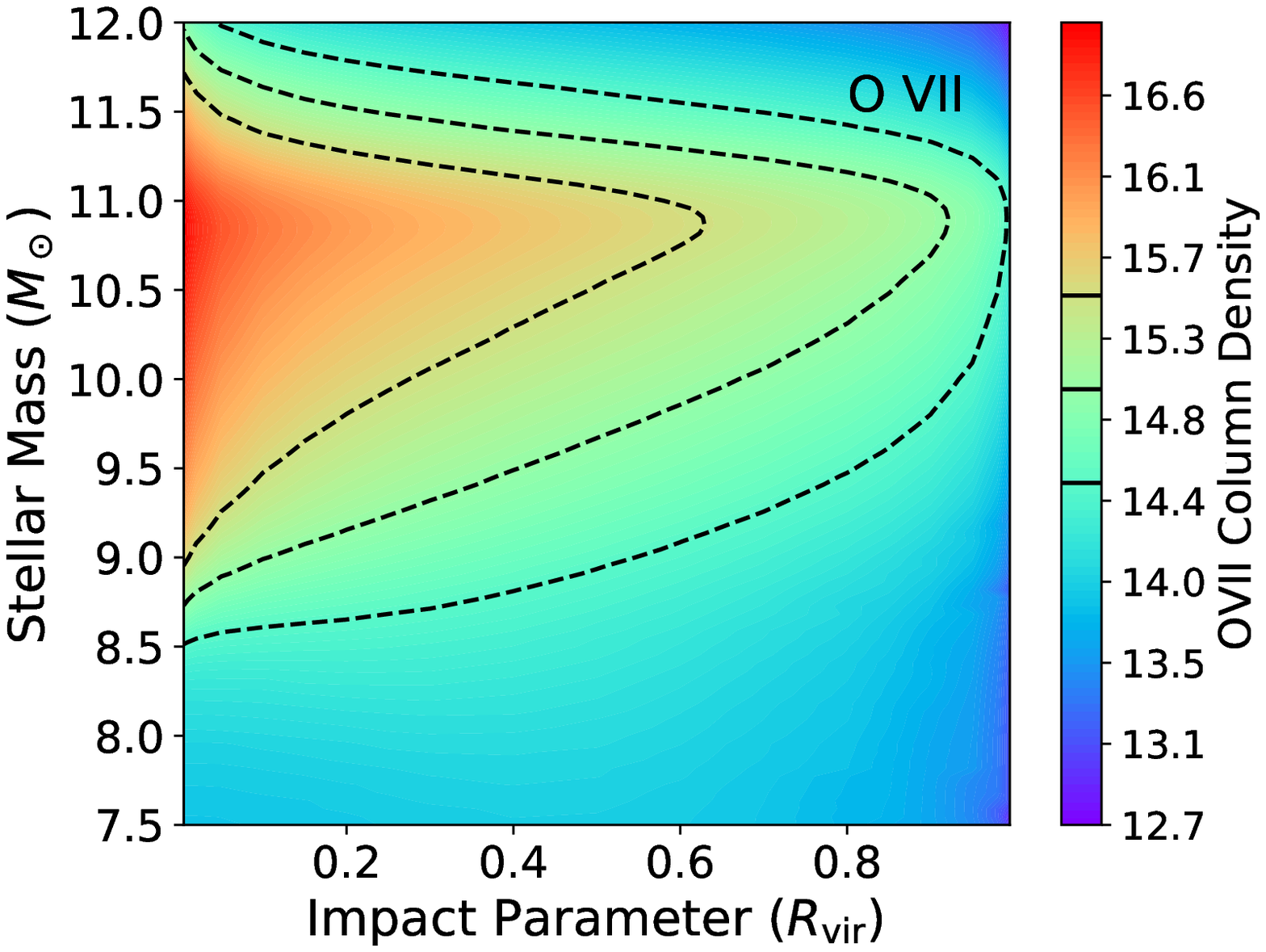}
\includegraphics[width=0.48\textwidth]{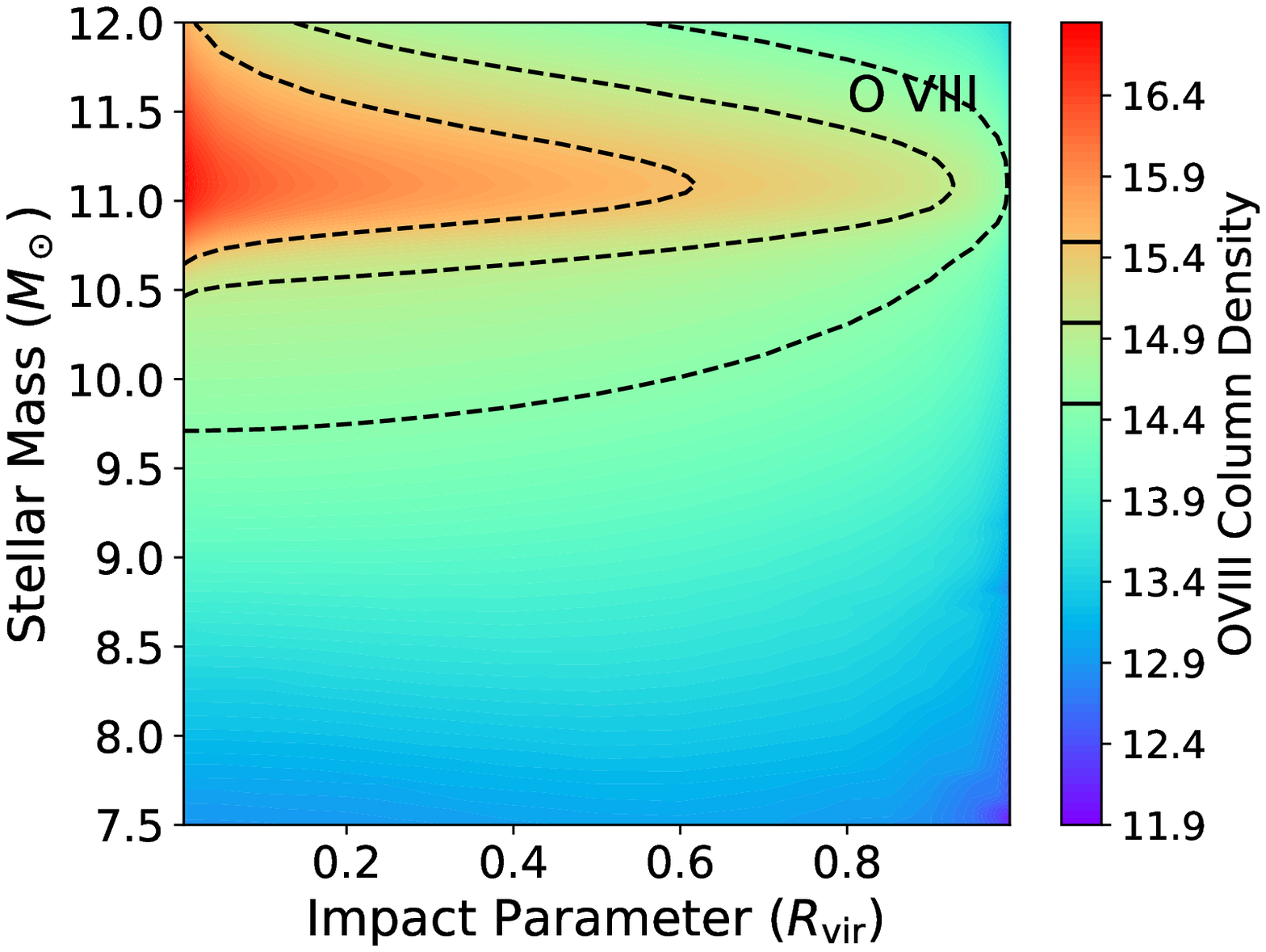}
}
\end{center}
\caption{The GGHM model for high ionization state ions (i.e., {\OVI}, {\NeVIII}, {\OVIII}, {\MgX}, {\OVIII}). The fiducial model has $Z=0.5Z_\odot$, $T_{\rm max}=2T_{\rm vir}$, $R_{\rm max}=R_{\rm vir}$ at $z=0$. For UV ions (i.e., {\OVI}, {\NeVIII}, {\MgX}; upper panels), the dashed contour lines are 13.5, 14.0 and 14.5 respectively. The X-ray ions (i.e., {\OVII} and {\OVIII}; lower panels), the contour levels are 14.5, 15.0 and 15.5, respectively.}
\label{halo_model}
\end{figure*}

\section{Methods}
To calculate the galaxy contribution to high ionization state ions in the universe, one needs a galactic gaseous halo model and the galaxy number density (i.e., the star-forming galaxy SMF). 
We adopted the gaseous halo model introduced in QB18, which connects the galaxy disk and the gaseous halo for star-forming galaxies.
Subsequently, the column density distribution is derived by convolving the gaseous halo model with the SMF.

\subsection{The Gaseous Halo Model}
The gaseous halo model is the {\tt TPIE} model in QB18, including the photoionization modification due to the ultraviolet background (UVB) and a time-independent model of the radiative-cooling multi-phase medium.
To summarize, our basic assumption is that the star formation rate (SFR) of the galaxy disk is balanced by the radiative cooling rate of the gaseous halo within the cooling radius, where $t_{\rm cooling}=t_{\rm Hubble}$. 
In QB18, we also consider heating due to feedback processes, which could modify the properties of the gaseous halo.
Specifically, we included a galactic wind model from the FIRE simulation \citep{Muratov:2015aa}, which is believed to have a tight relationship with the SFR. 
In our modeling, a $\gamma$ factor (${\rm SFR} = \gamma \dot{M}_{\rm cooling}$) is introduced to include the heating by stellar feedback, which varies from $0.14$ to $0.70$ from low-mass galaxies ($M_\star=3\times 10^7~M_\odot$) to massive galaxies ($M_\star = 10^{12}~M_\odot$).
Therefore, once the SFR of the galaxy is obtained, the radiative cooling rate is determined, which will be used to calculate the density of the gaseous halo.
The SFR is fixed for different stellar masses and redshifts using the star formation main sequence \citep{Pannella:2009aa, Morselli:2016aa}. 
Generally, the low-mass galaxies have lower SFR values but higher specific SFR (${\rm sSFR} = {\rm SFR}/M_\star$) and higher redshift galaxies have higher SFR, roughly $\overline{\rm SFR} \propto (1+z)^3$.

Our multi-phase medium model assumes that there is a hot ambient gas supplying the ``steady-state" gaseous halo and the mass cooling rate is constant over different temperatures, which lead a universal distribution for the mass-temperature distribution (QB18). 
Photoionization modifies the cooling curve and the ionization fractions at different temperatures; ionization equilibrium is assumed.
The gas density profile is fixed as a $\beta$-model ($\rho(r) = \rho_0 r^{-3\beta}$) and $\beta$ is fixed to 0.5 such that the halo is near hydrostatic equilibrium. 
In the GGHM model, there are also free parameters -- the upper temperature limit ($T_{\rm max}$; the ambient gas temperature), the lower temperature limit ($T_{\rm min}$; setting the cooling temperature range), the metallicity ($Z$) and the maximum radius ($R_{\rm max}$; setting the cutoff of the $\beta$-model). 
Since we only consider the high ionization state ions, our results are not sensitive to the lower temperature limit, therefore, it is fixed to the fiducial value ($3\times10^4\rm~K$) in QB18. 

We explore the effect of varying the other parameters ($T_{\rm max}$, $Z$, and $R_{\rm max}$) in Section \ref{results}, but here present a fiducial model.
We adopt twice the virial temperature as the fiducial $T_{\text{max}}$.
QB18 showed that a gaseous halo at the virial temperature significantly underestimates the total amount of {\OVIII} gases of the MW.
The preferred model indicated that the MW has an ambient gas temperature about twice the virial temperature.
This is consistent with X-ray observations of external galaxies, which shows that the gas temperature is about twice the virial temperature \citep{Goulding:2016aa}.
In the same model, the derived metallicity of the MW gaseous halo is $0.5-0.6~Z_\odot$ (as argued by \citealt{Faerman:2017aa} and \citealt{Bregman:2018aa}).
The column density measured around the MW is not sensitive to $R_{\rm max}$ in the $\beta$-model.
Raising the maximum radius from the virial radius to twice the virial radius only increases the {\OVII} and {\OVIII} column densities by $8\%$ (within the measurement errorbar $\approx 20\%$).
Therefore, the maximum radius is unconstrained by Galactic observations, but a larger halo does significantly increase the detectable cross-section (Section \ref{results}).
Thus, our fiducial model has $T_{\rm{max}} = 2T_{\rm{vir}}$, $Z = 0.5 Z_{\odot}$, and $R_{\rm{max}} = R_{\rm{vir}}$. 

In Fig. \ref{halo_model}, we show the high ionization state ion column densities in the fiducial model, quantifying the dependence on the stellar mass and the impact parameter.
In the GGHM model, the origins of high ionization state ions (i.e., {\OVI}, {\NeVIII}, {\OVII}, {\MgX}, {\OVIII}) could be divided into three categories: the photoionized virialized gaseous halo; the collisional virialized gaseous halo; and the radiatively-cooling flow (QB18).
As galaxy mass increases, collisional ionization becomes more important than photoionization, with the transition at the mass where the virial temperature corresponds to the peak ionization fraction for each ion.
Each ion also has the highest column density around the galaxies with the transition mass, which is set by the ionization potential.
The transition galaxy mass for {\OVI} is around $M_\star \approx 3\times 10^9 ~ M_\odot$ or halo masses of  $M_{\rm h} \approx 2\times 10^{11} ~ M_\odot$, while the transition stellar masses are $1\times 10^{10}~M_\odot$, $4\times10^{10}~M_\odot$, $8\times10^{10}~M_\odot$, and $1\times 10^{11}~M_\odot$ for {\NeVIII}, {\OVII}, {\MgX}, and {\OVIII}, respectively.
Above the transition mass, the ions are generated in the cooling flows from the hotter ambient gas.
The radial dependence of the column density decreases as one expects from the $\beta$-model since the ionization fraction is constant over different radii.
Below the transition mass, photoionization becomes more important, which weakens the column density dependence on the radius (showing a flattened distributions in the inner region) and allows high ionization state ions to exist for a wide range of stellar masses (see QB18 for details).

\begin{figure*}
\begin{center}
\subfigure{
\includegraphics[width=0.48\textwidth]{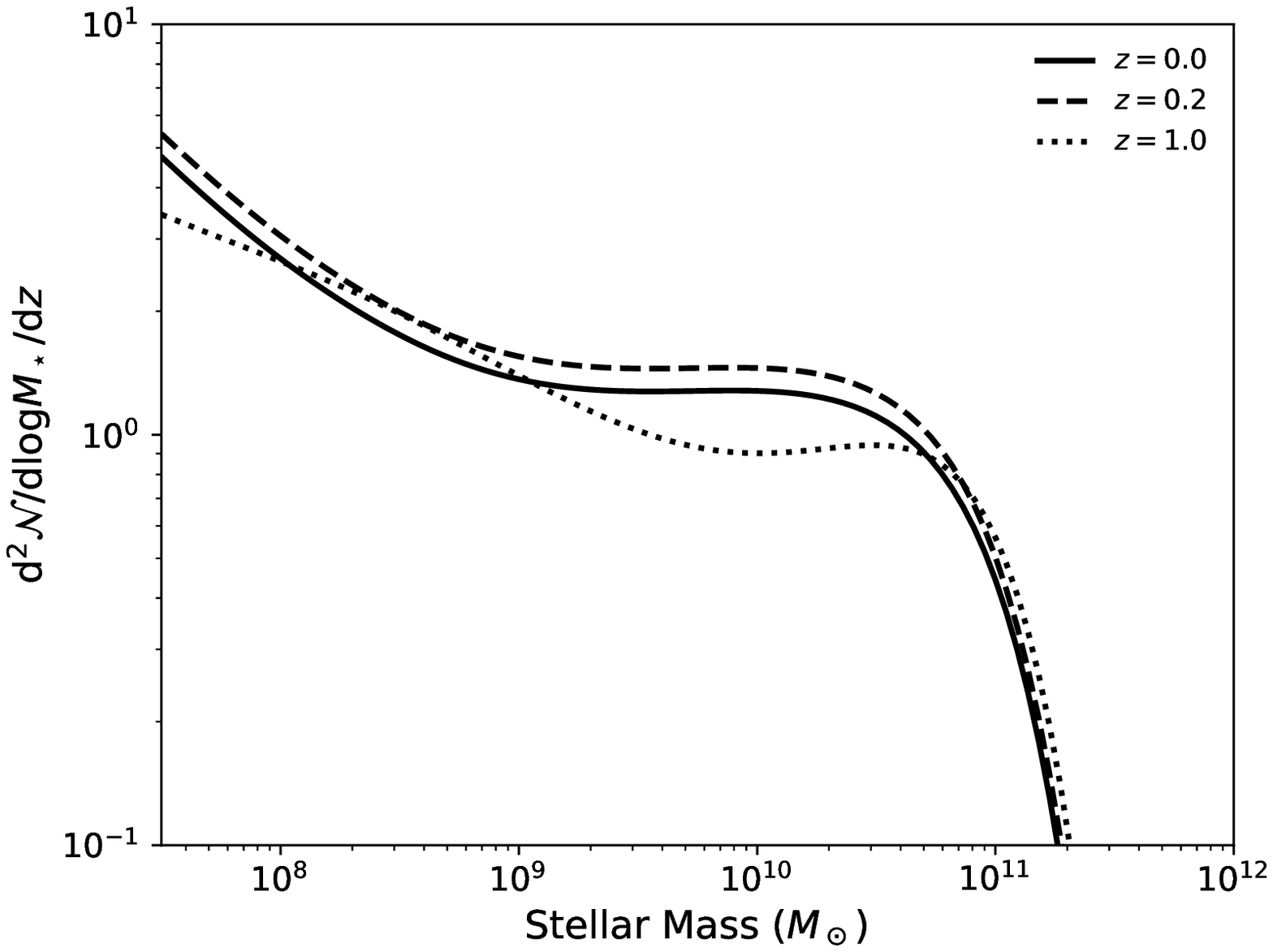}
\includegraphics[width=0.48\textwidth]{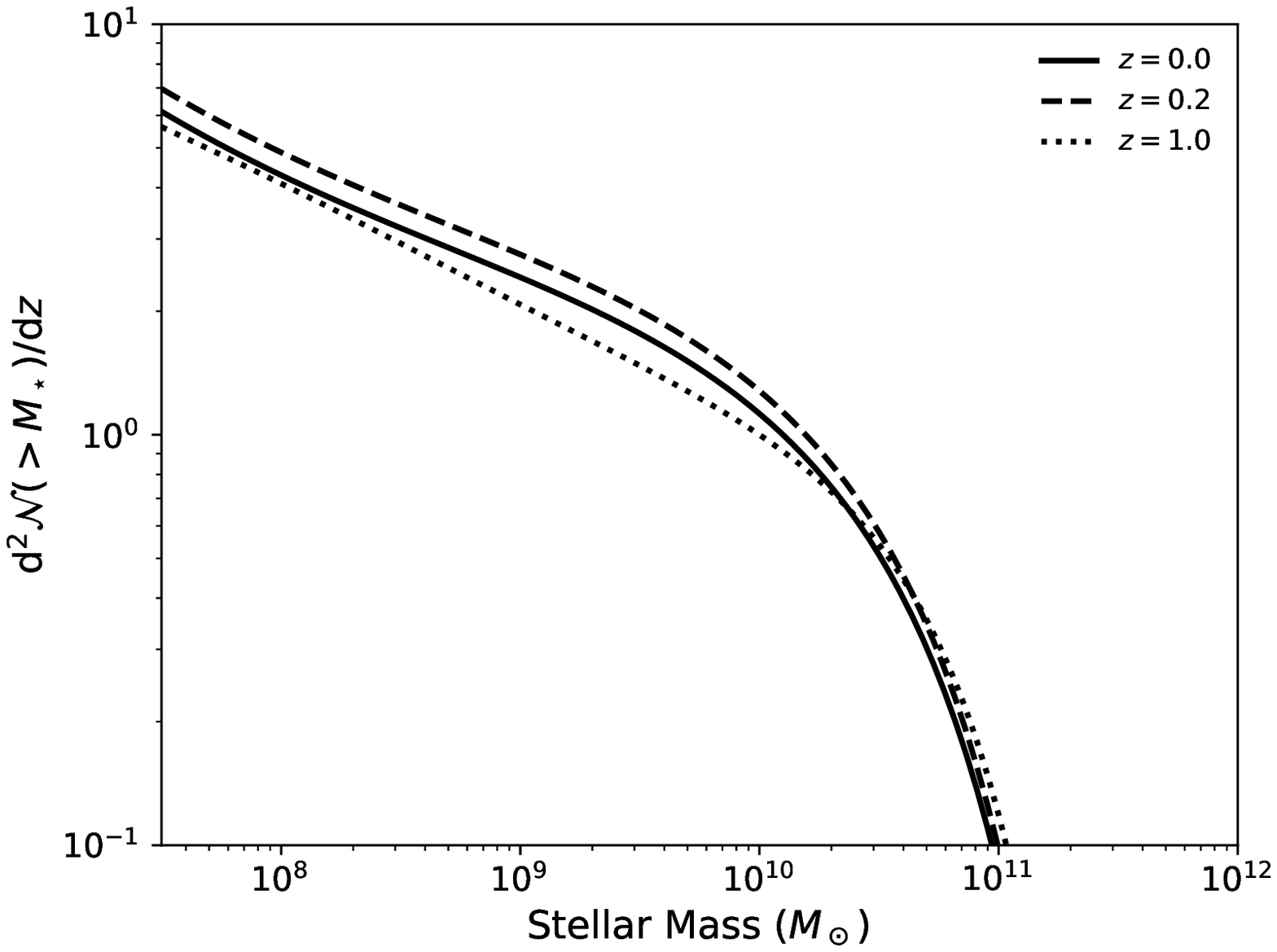}
}
\end{center}
\caption{{\it Left panel}: The differential detection rate of gaseous halos as a function of galaxy stellar mass assuming the maximum radius is the virial radius. The adopted SMF is only for star-forming galaxies \citep{Tomczak:2014aa}. \red{{\it Right panel}: The cumulative detection rate of gaseous halos.}}
\label{gh_cf}
\end{figure*}

\begin{figure*}
\begin{center}
\subfigure{
\includegraphics[width=0.48\textwidth]{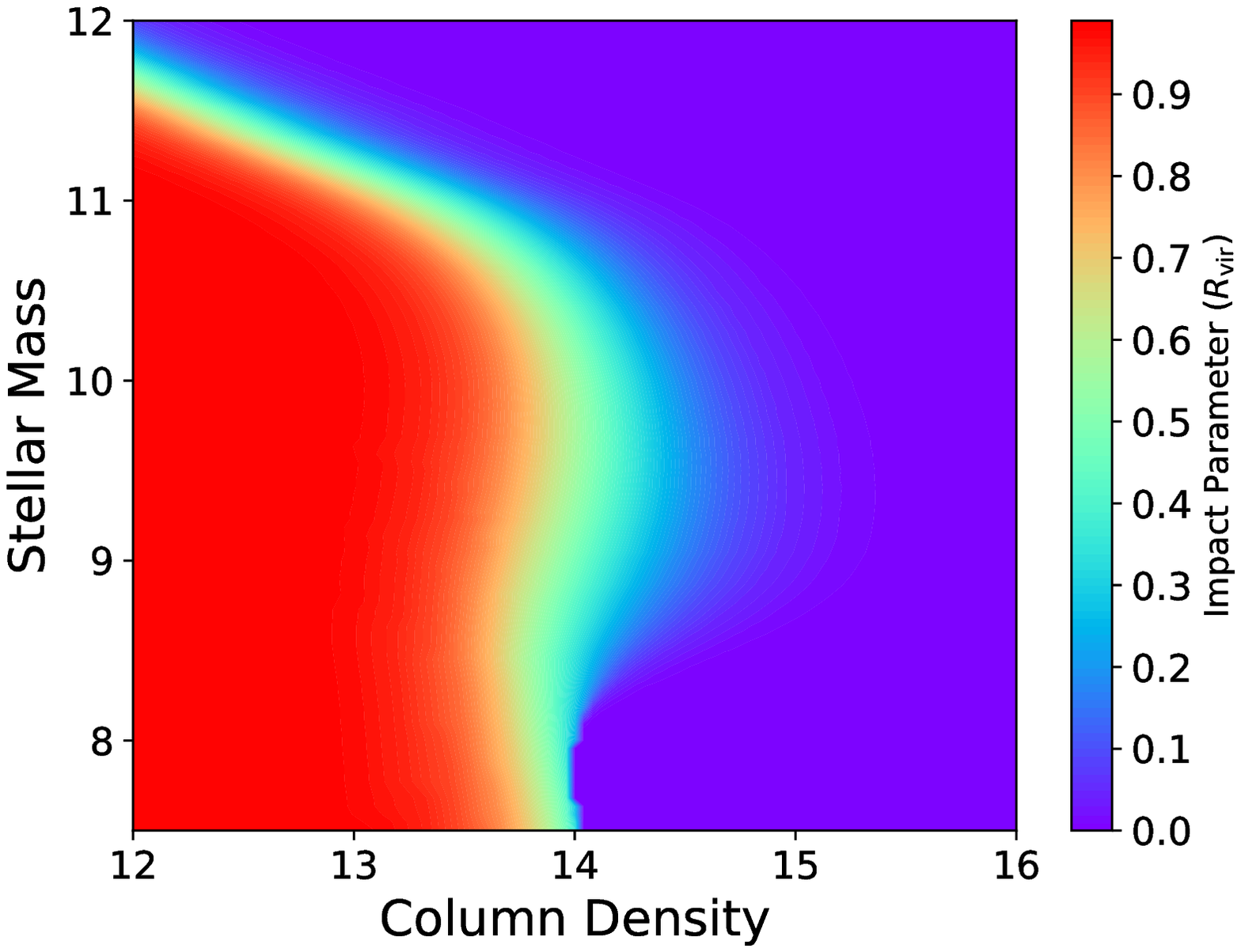}
\includegraphics[width=0.48\textwidth]{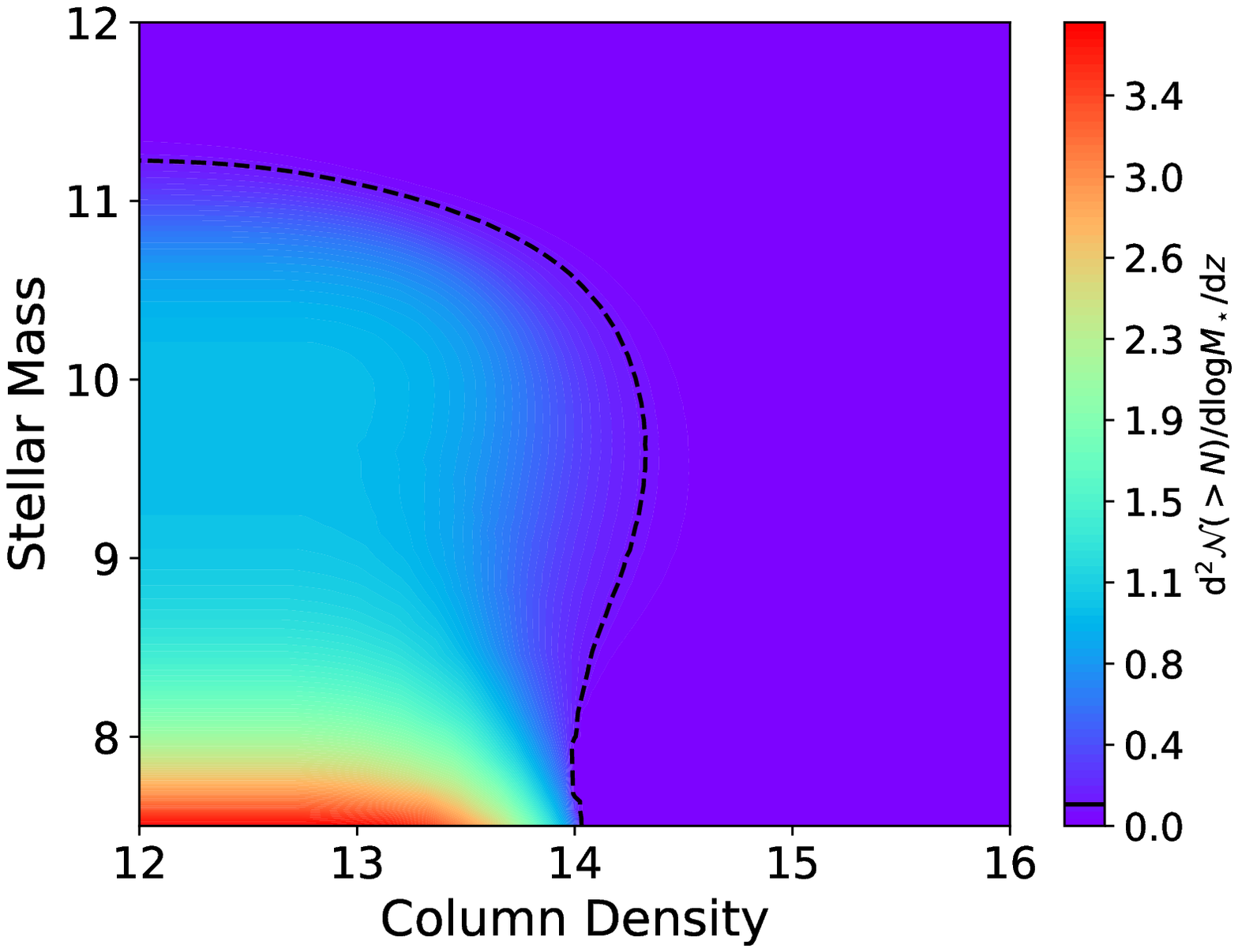}
}
\end{center}
\caption{{\it Left panel}: The relative impact parameter of {\OVI} at different column densities. Within the relative impact parameter, the {\OVI} column density is larger than the given value as marked by the x-axis. {\it Right panel}: The cumulative {\OVI} column density distribution at different stellar masses. The black dashed line shows the contour of a detection rate of $0.1~\rm dex^{-1}$ per unit redshift, which indicates sub-$L^*$ galaxies are the major contributors to the high {\OVI} column density systems.}
\label{abund_model}
\end{figure*}

\subsection{The Contribution of Galaxies}
To consider the cosmic galaxy contribution, we adopt the SMF to represent the galaxy number density in the universe.
The adopted redshift-dependent SMF is from \citet{Tomczak:2014aa}, where the SMF is calculated in redshift bins of $\Delta z \approx 0.2$. \
In Fig \ref{gh_cf}, we show the path-length density (detection rate) of galactic gaseous halos at different redshifts where the maximum radius equals the virial radius.
\red{Since the difference between the SMF at $z=0.0$ and $z=0.5$ is small ($< 10\%$; \citealt{Behroozi:2013aa}), we use the same SMF at $0.2< z<0.5$ for these two redshifts (\citealt{Tomczak:2014aa}; in $\log M_\star = 8.00 - 11.25$).
At $z=1.0$, we use the SMF for the redshift bin of $z=0.75- 1$, which is measured in the stellar mass range of $\log M_\star = 8.50-11.25$.}
As the redshift increases, the comoving galaxy density decreases, while the comoving covering area per gaseous halo increases due to increasing of the scale factor.
Overall, the detection rates at different redshifts are similar, showing total cross section values of ${\rm d}\mathcal{N}/{\rm d}z = 6.0$, 6.8 and 5.5 at $z=0.0$, $0.2$, and $1.0$, respectively.
However, the similarity of total detection rate does not imply that the column density distribution is also similar to each other.

For a given column density, only a part of the gaseous halo cross section contributes to the detection of systems that are higher than the column density.
To represent the cross sections of different column densities, we define the relative impact parameter (or radius), within which the column density due to the gaseous halo is higher than the given column density (the left panel of Fig. \ref{abund_model} for {\OVI}).
Since the inner regions typically have higher column densities, this treatment applies in most stellar mass ranges for all ions; while for the most low-mass galaxies, there is a modest decrease in the inner region (of 0.2 dex) because of the photoionization modification (QB18).
However, one expects that there are ionizing photons escaping from the galaxy disk, which will boost the high ionization state ion column density in the inner region, although the treatment of an escaping flux contribution is beyond the scope of this paper.
Therefore, we ignore the decrease in the innermost regions of low-mass galaxies and apply the relative impact parameter to all galaxies and all column densities.

Applying the relative impact parameter ($r_{\rm cf}$), we calculate the cumulative detection rates at different column densities for different stellar masses:
\begin{equation}
\frac{{\rm d}^2 \mathcal{N}(>N)}{{\rm d} z {\rm d} M_{\star}} = \pi (r_{\rm cf} R_{\rm vir} (1+z))^2 \times \mathcal{F}(M_\star),
\end{equation}
where $z$ is the redshift to account for the scale factor and $\mathcal{F}(M_\star)$ is the comoving SMF at different stellar masses. 
Our results are shown in the right panel of Fig. \ref{abund_model} for {\OVI}, showing that the high {\OVI} column density ($N>10^{14}\rm~cm^{-2}$) systems are mainly from the galaxies with stellar masses between $10^{8.5}~M_\odot$ and $10^{10.5}~M_\odot$, while low column density systems occur in all galaxies.
The cumulative column density distribution is calculated by integrating over all stellar masses:
\begin{equation}
\mathcal{F}(N) = \frac{{\rm d}\mathcal{N}(>N)}{{\rm d}z } = \int \frac{{\rm d}^2 \mathcal{N}(>N)}{{\rm d} z {\rm d} M_{\star}} {\rm d} M_\star,
\end{equation}
where $\mathcal{F}(N)$ is the cumulative detection rate for different column densities. Meanwhile the column density distribution function is defined as
\begin{equation}
f(N) = \frac{{\rm d}\mathcal{N}(>N)}{{\rm d}z {\rm d} N}  = \frac{{\rm d} \mathcal{F}(N) }{{\rm d} N}.
\end{equation}
In the following sections, we mainly use the cumulative column density distribution for comparison with observations, while the column density distribution function is used to compare with cosmological simulations.

\begin{figure*}
\begin{center}
\subfigure{
\includegraphics[width=0.48\textwidth]{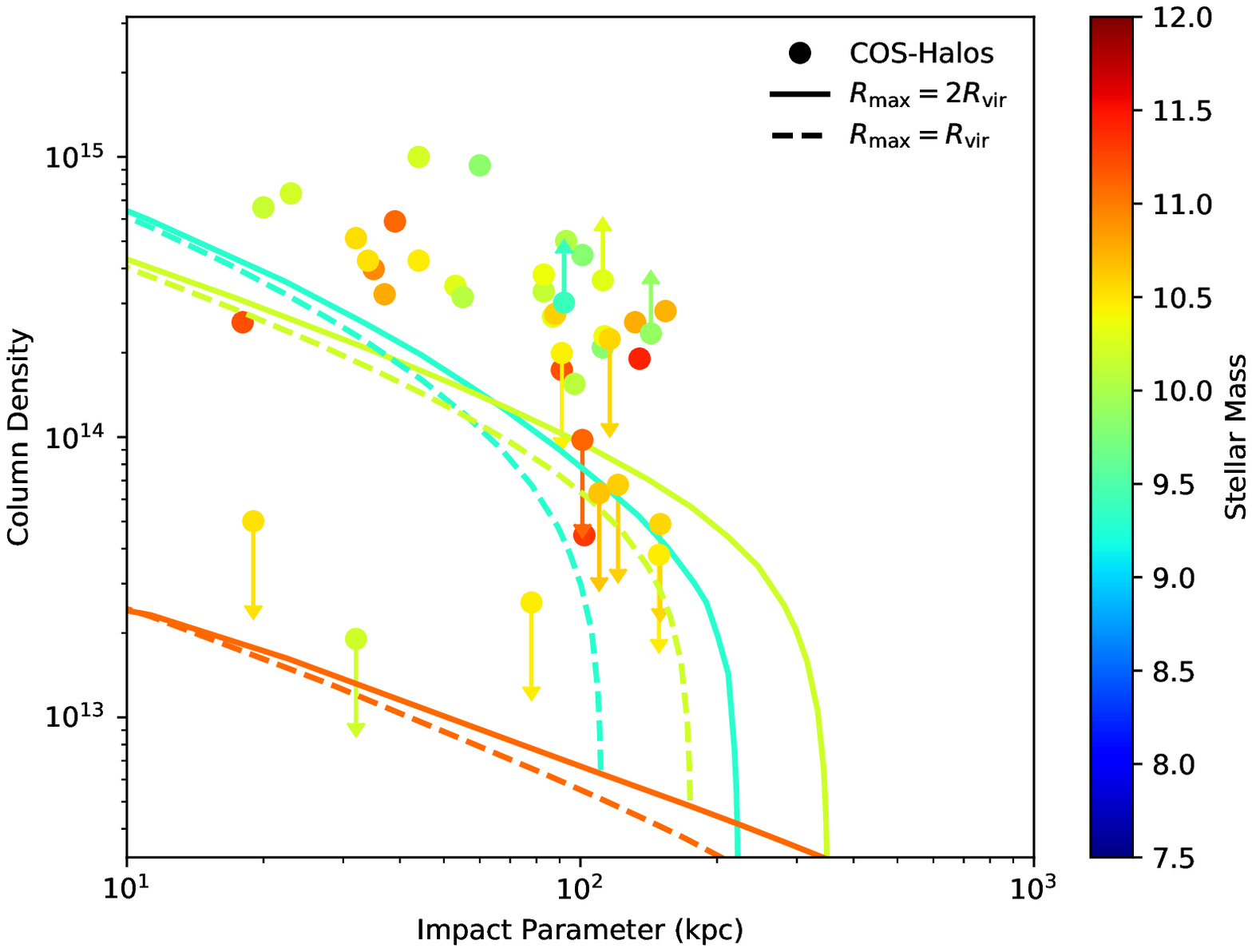}
\includegraphics[width=0.48\textwidth]{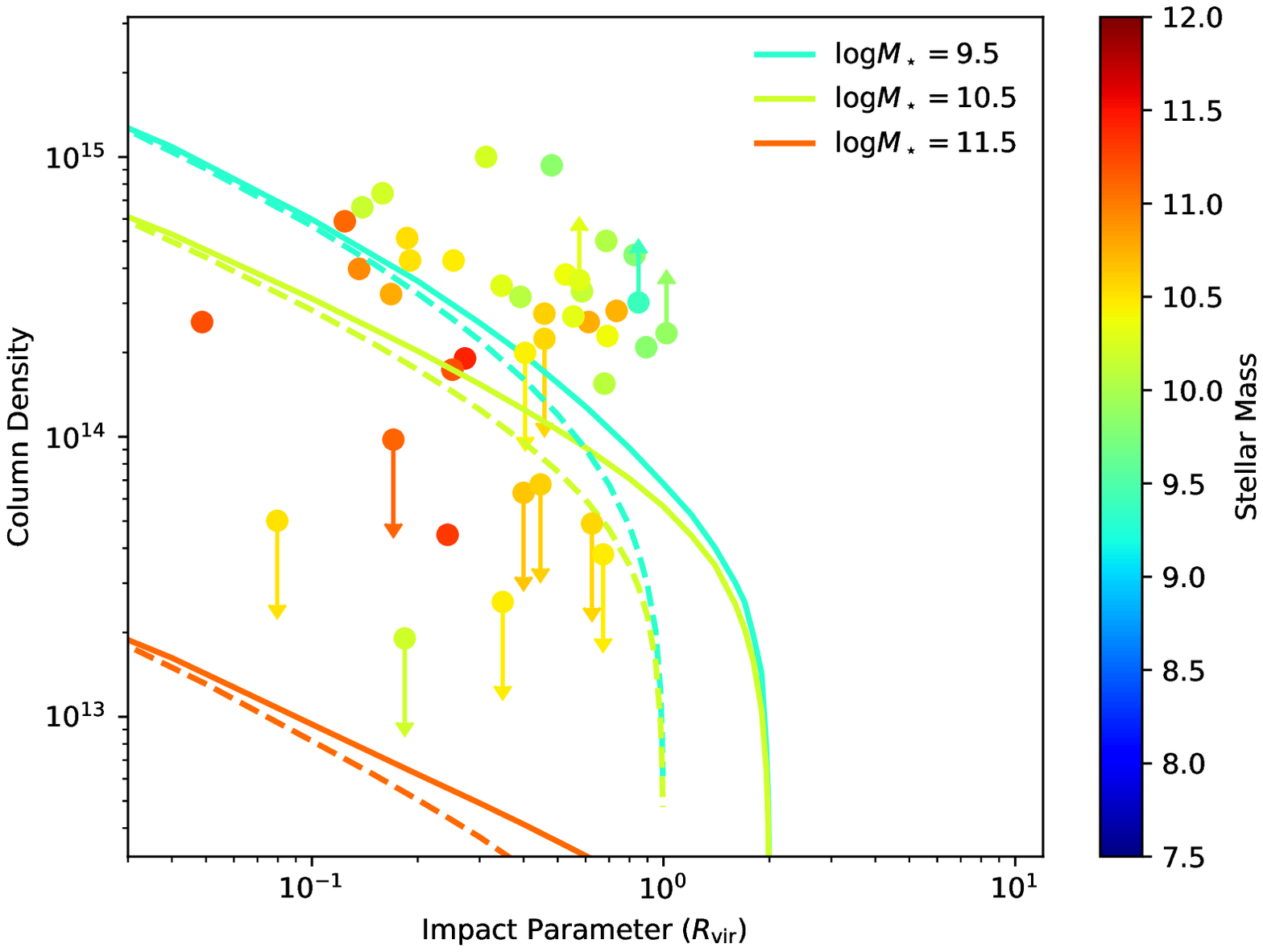}
}
\subfigure{
\includegraphics[width=0.48\textwidth]{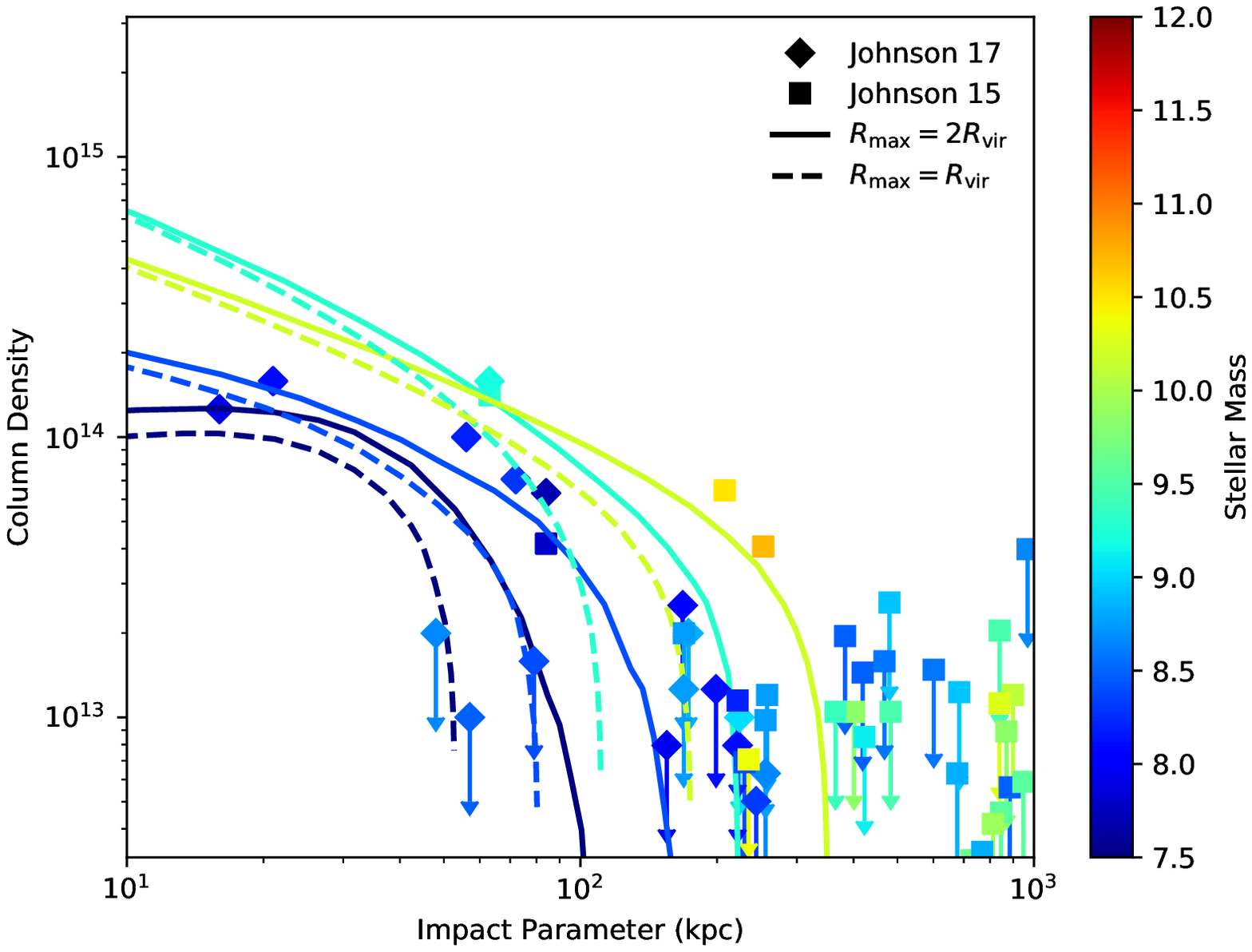}
\includegraphics[width=0.48\textwidth]{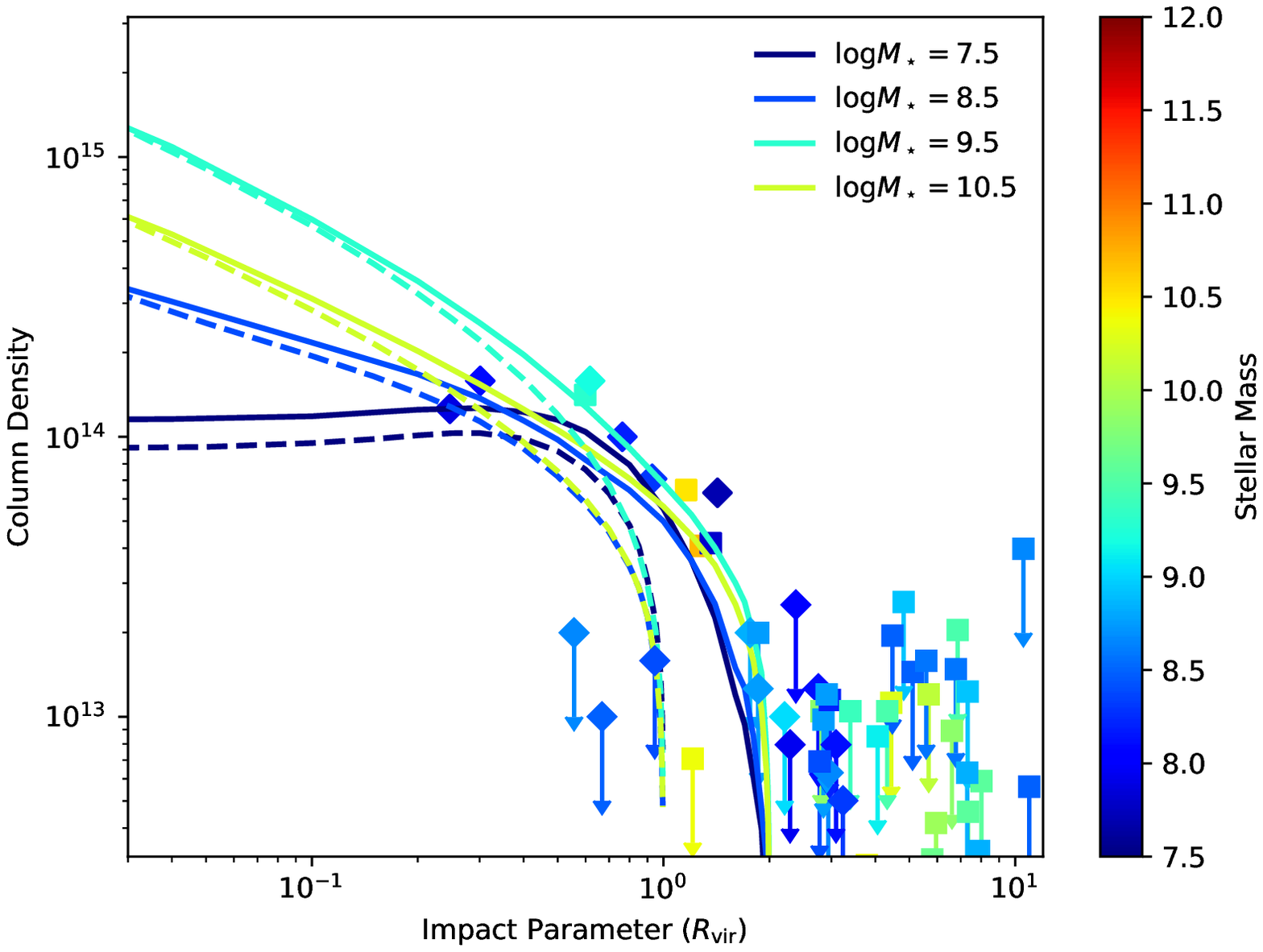}
}
\end{center}
\caption{The radial dependence of {\OVI} in the GGHM models and observations. The GGHM models are for galaxies with stellar masses of $\log M_\star = 7.5$, 8.5, 9.5, 10.5, and 11.5 at $z=0.2$, respectively. The observations are from COS-Halos (\citealt{Werk:2014aa}; marked as circles; upper panels), and the surveys of (\citealt{Johnson:2015aa}; square; lower panels), and (\citealt{Johnson:2017aa}; diamond; lower panels). Both the models and the observations are color-coded by the stellar masses. {\it Left panels}: Dependence on the physical impact parameter. {\it Right panels}: Dependence on the impact parameter in the units of the virial radius. Low mass galaxies ($\log M_\star < 8.5$) have a flattened radial dependence, which is consistent with Johnson's sample. Higher mass galaxies ($\log M_\star > 8.5$) show a decline with radius, which is consistent with the COS-Halos sample, but the {\OVI} column density from COS-Halos is systematically higher than our model predictions.}
\label{OVI_imp}
\end{figure*}

\begin{figure*}
\begin{center}
\subfigure{
\includegraphics[width=0.32\textwidth]{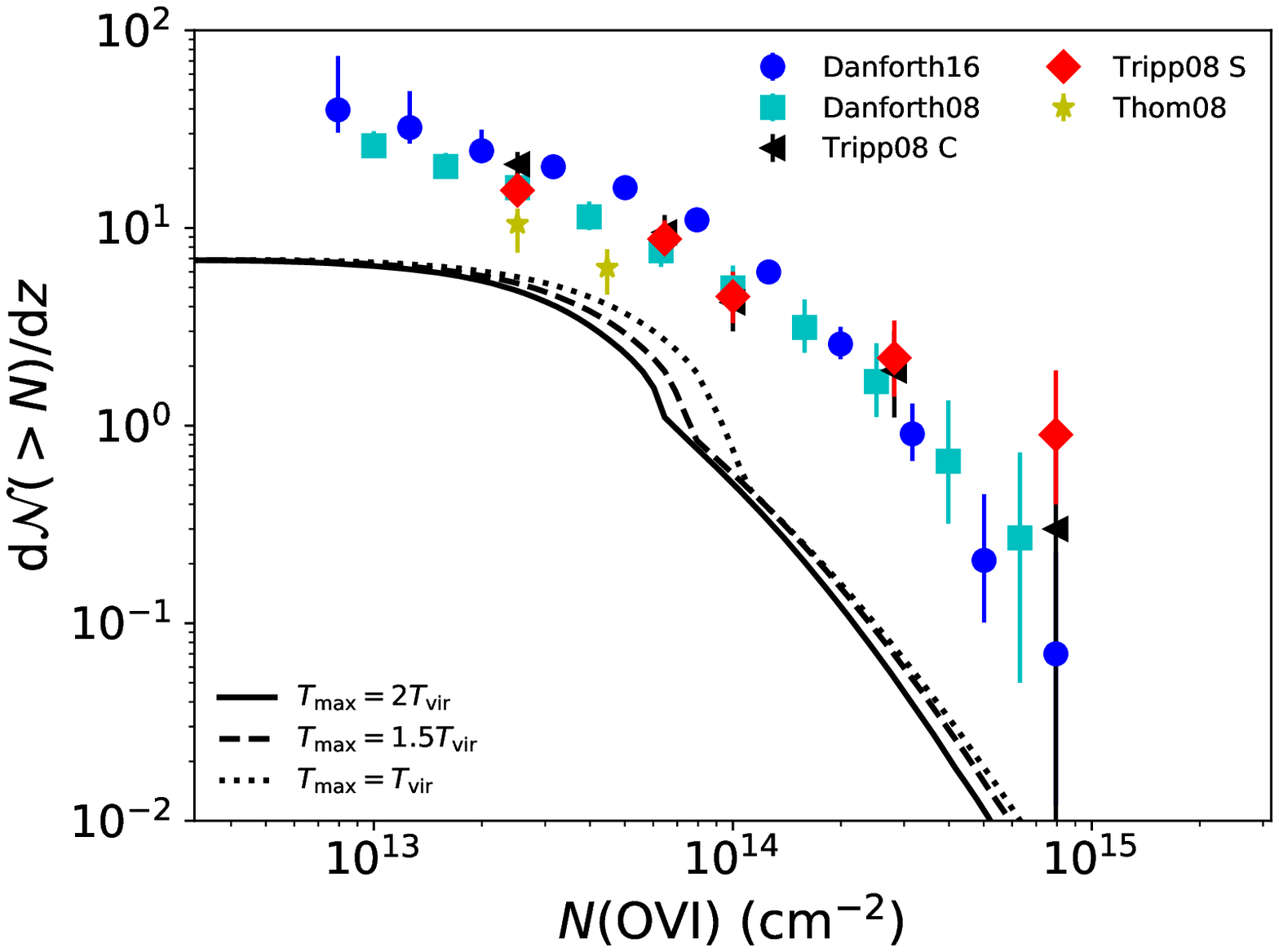}
\includegraphics[width=0.32\textwidth]{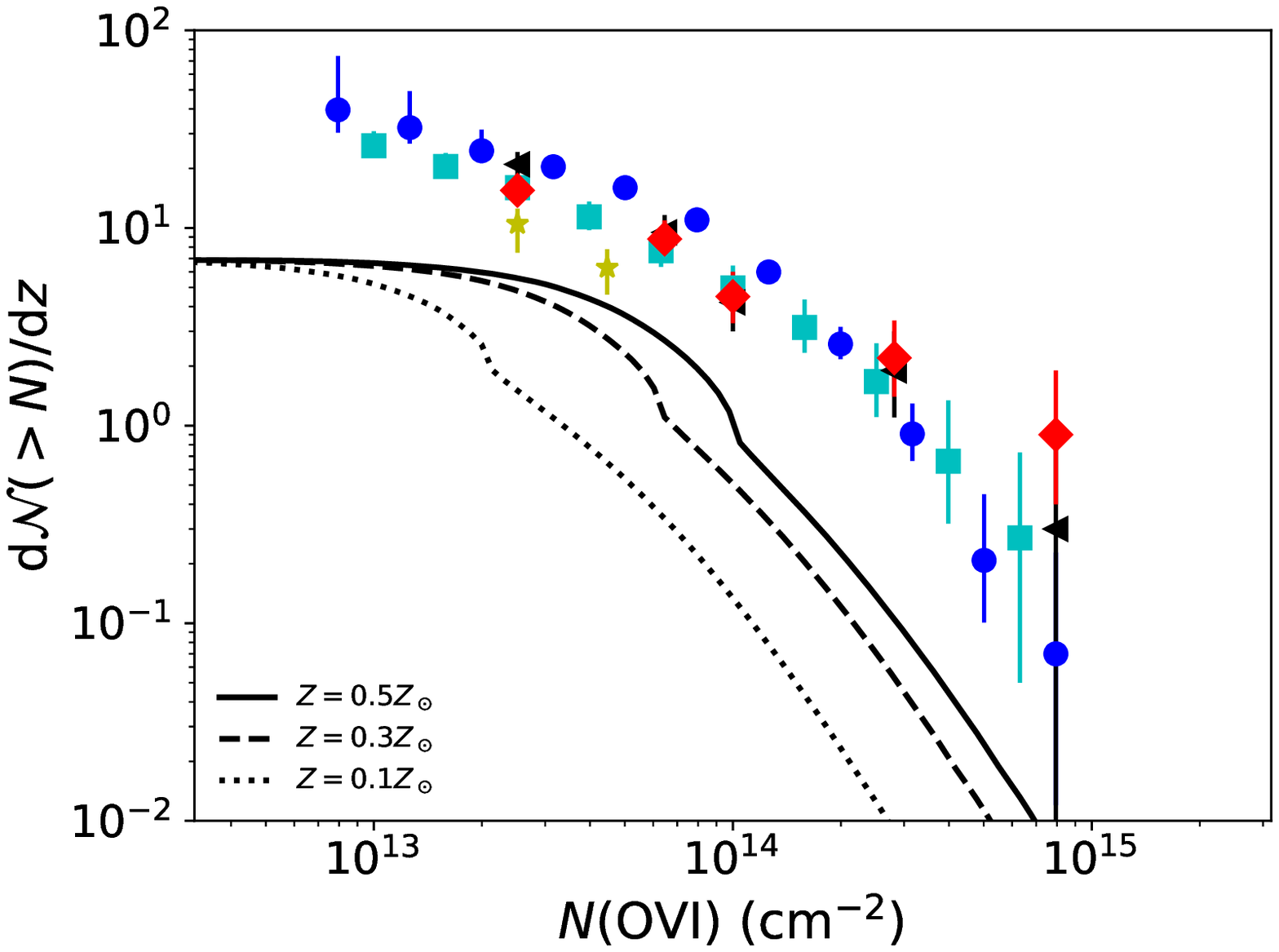}
\includegraphics[width=0.32\textwidth]{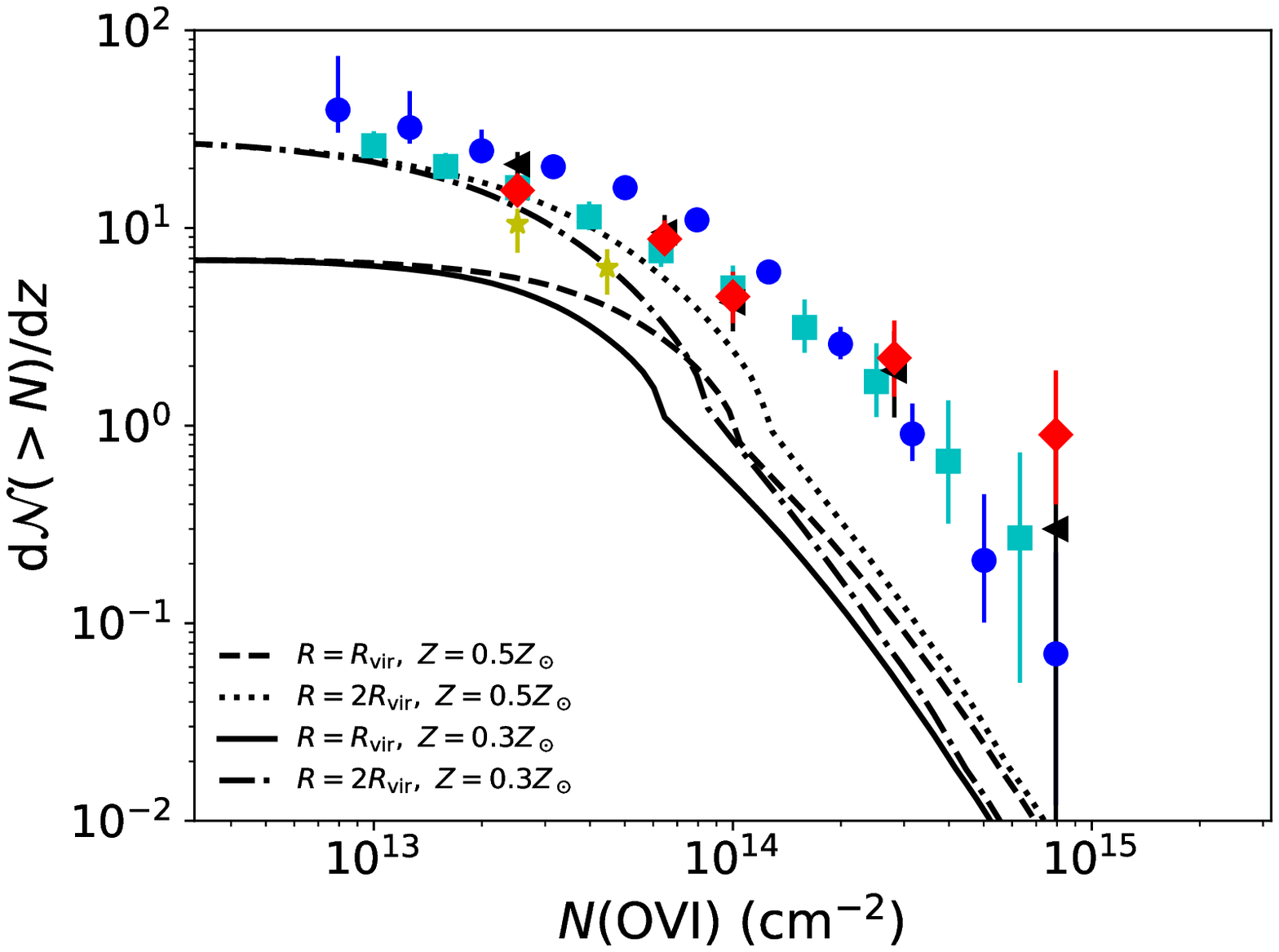}
}
\end{center}
\caption{The predicted column density distributions in the GGHM models with varied parameters at $z=0.2$. The data consist of intervening {\OVI} with or without galaxy information. The data are from \citet{Danforth:2008aa}, \citet{Thom:2008aa}, \citet{Tripp:2008aa}, and \citet{Danforth:2016aa}. For \citet{Tripp:2008aa}, two column density distributions are shown for components (``C") or systems (``S"). In each panel, we show the result of varying one parameter while keeping the others fixed at the fiducial values ($T_{\rm max}=2T_{\rm vir}$, $Z=0.5 Z_\odot$, and $R_{\rm max}=R_{\rm vir}$). {\it Left panel}: The dependence on $T_{\rm max}$. {\it Middle panel}: The dependence on $Z$. {\it Right panel}: The dependence on $R_{\rm max}$.}
\label{TZR}
\end{figure*}

\section{The High Ionization State Ions}
\label{results}
Once the ion column density distribution is calculated, we can compare the model with observations to constrain the free parameters and the physical conditions of observed systems. For {\OVI} systems, there are a variety of observations to be compared with, while for other ions, we mainly show the predictions of the galaxy contributions.

\subsection{Intervening {\OVI} at $z \approx 0.2$}
Observations constrain the column density as a function of impact parameter.
In Fig. \ref{OVI_imp}, three data sets are shown -- the COS-Halos sample \citep{Werk:2013aa, Werk:2014aa}, the \citet{Johnson:2015aa} sample, and a dwarf galaxy sample \citep{Johnson:2017aa}. 
For the comparison models, we adopt $z=0.2$ with $R_{\rm max} = R_{\rm vir}$ or  $R_{\rm max} = 2R_{\rm vir}$, since each sample has a median redshift $z\approx0.2$.
The COS-Halos sample selected $L^*$ galaxies ($\log M_\star = 9.7-11.5$) within the virial radius, while \citet{Johnson:2017aa} focus on dwarf galaxies ($\log M_\star = 7.7-9.2$).
The \citet{Johnson:2015aa} work shows a QSO-selected sample with impact parameters up to ten times the virial radius, which probes beyond the halo itself, and we selected isolated and late-type galaxies from this sample. 
\citet{Savage:2014aa} presented an {\OVI} absorption-based sample of galaxy-absorption pairs, which is biased to detect {\OVI} absorption features not near known galaxies; therefore, we do not include this sample in the comparison (see discussion in Section \ref{igm}). 
\red{In the comparison with the COS-Halos, our models underestimate the observed \ion{O}{6} column densities by a factor of $\approx 0.3\rm~ dex$ systematically, which is discussed in detail in QB18.
Nevertheless, the radial decline of the $\approx L^*$ galaxies ($M_\star \gtrsim 10^{10}~M_\odot$) is reproduced with the similar magnitude.
The maximum \ion{O}{6} column density systems occur in around $M_\star \approx 10^{9.5} - 10^{10.5} ~M_\odot$.
For low mass galaxies ($M_\star \lesssim 10^{8.5}~M_\odot$), our models appear to match the data showing the turnover in the radial direction (the flattened radial distribution; \citealt{Johnson:2017aa}).}
Also, the modeling favors the solution with the $R_{\rm max} = 2R_{\rm vir}$ to best reproduce the observations based on measurements and non-detections of the \citet{Johnson:2015aa} sample.

\red{We also consider {\OVI} absorption blind survey samples that are different from galaxy-QSO pair samples.}
In Fig. \ref{TZR}, we show the comparison with these four samples -- \red{\citet{Danforth:2008aa}, \citet{Thom:2008ab},\citet{Tripp:2008aa}, and \citet{Danforth:2016aa}, which have median redshifts of $0.20$, $0.25$, $0.22$, and $0.29$.}
In these four samples, the detected {\OVI} systems may not have galaxy information and all of the detected  {\OVI} contributes to the total column density distribution.
Specifically, these samples have redshift regions of \red{$z < 0.36$, $0.1<z<0.5$, $z < 0.5$, $0.1< z<0.7$, respectively.}
For \citet{Danforth:2008aa} and \citet{Danforth:2016aa}, we used the reported differential column density distribution to calculate the cumulative column density distribution.
For \citet{Tripp:2008aa}, we assume a Doppler $b$ factor of $30\rm~\kms$ to convert the rest-frame equivalent width (EW) into a column density, since the average $b$ factor is $27\pm 14 \kms$ in this sample. 
In \citet{Tripp:2008aa}, there are two reported column density distributions based on whether a system is broken into separate components -- components (``C") and systems (``S").
Therefore, the ``S" column density distribution is more flattened, having more high column systems.
Our fiducial model has the parameters $Z=0.5~Z_\odot$, $T_{\rm max} = 2~T_{\rm vir}$, and $R_{\rm max} = R_{\rm vir}$ at $z=0.2$. 
The results of varied $Z$, $T_{\rm max}$ and $R_{\rm max}$ are also shown in Fig. \ref{TZR}.

All of the GGHM models underestimate the detection rate of the observed {\OVI}, showing a gap with a factor of $\approx 4-10$ over all column densities (Fig. \ref{TZR}).
However, the GGHM models predict the general shape of the observed column density distribution.
\citet{Danforth:2016aa} shows that the column density distribution can be fitted by a broken power law with a break point at $\log N({\rm OVI}) = 14.0\pm 0.1$ and two power law indices of $2.5 \pm 0.2$ (high column density end) and $0.56 \pm 0.16$ (low column density end).
Our models show a similar break around $\log N({\rm OVI})=14$, which is indicated by a relatively sharp decrease of the detection rate.
As we will show in Section 4, this decrease is due to the contribution from low-mass galaxies and the break indicates that there is a lower column density limit for {\OVI} systems with galaxy origins.
As shown in Fig. \ref{halo_model}, the majority of the gaseous halos have $N({\rm OVI})\approx 10^{14}\rm~cm^{-2}$ for low-mass galaxies ($\lesssim 10^{8.5}~M_\odot$), which leads to the sharp decrease of the cumulative column density distribution around the break point.

Increasing the maximum temperature moves the {\OVI} transition galaxy mass from $\log M_{\star} = 10.2$ ($T_{\rm max} = T_{\rm vir}$) to $9.5$ ($T_{\rm max} = 2T_{\rm vir}$) and affects the column density distribution by changing the detection rate of the corresponding galaxies.
As shown in Fig. \ref{gh_cf}, the detection rate of galaxies changes only modestly from $10^{10}~M_\odot$ ($T_{\rm max} = T_{\rm vir}$) to $10^{9.5}~M_\odot$ ($T_{\rm max} = 2 T_{\rm vir}$), which corresponds to high column density {\OVI} systems.
Therefore, the high column density end of the {\OVI} column density distribution does not change significantly.
Higher ionization state ions (e.g., {\OVII} and {\OVIII}) could be affected significantly, since there are many fewer massive galaxies.
We also notice that the break point is slightly smaller for the high temperature model.
This is mainly because the high temperature model predicts lower column densities in the low-mass galaxies that contribute to the break.
In low-mass galaxies, the radiative cooling is suppressed by photoionization, and increasing the temperature reduces the impact of photoionization (see Fig 1. in QB18).
The high radiative cooling emissivity could reduce the gas density in high temperature gaseous halo models, which leads to lower {\OVI} column densities.

Increasing the metallicity helps with the problem that our models lie below the observation, since raising the metallicity increases the {\OVI} column density in all galaxies (QB18). 
This effect moves all column density toward higher values, which can make up some of the gap at the high column density end (shown in the middle panel of Fig. \ref{TZR}). 
However, varying only the metallicity cannot make up the entire gap between the model and observations, since it cannot increase the total detection rate of {\OVI} in the low-column density range.

\begin{figure}
\begin{center}
\includegraphics[width=0.48\textwidth]{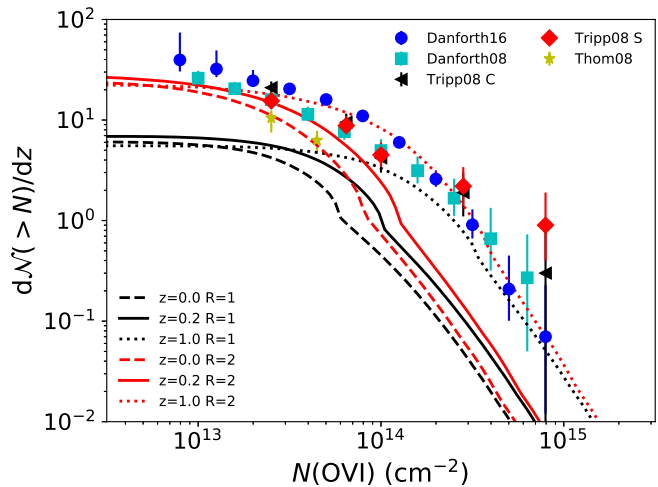}
\end{center}
\caption{The redshift evolution of the {\OVI} column density distribution. The observations used are color-coded as Fig. \ref{TZR}. The higher redshift leads to the higher specific SFR, which raises the {\OVI} columns of all galaxies.}
\label{OVI_z}
\end{figure}

\begin{figure*}
\begin{center}
\subfigure{
\includegraphics[width=0.48\textwidth]{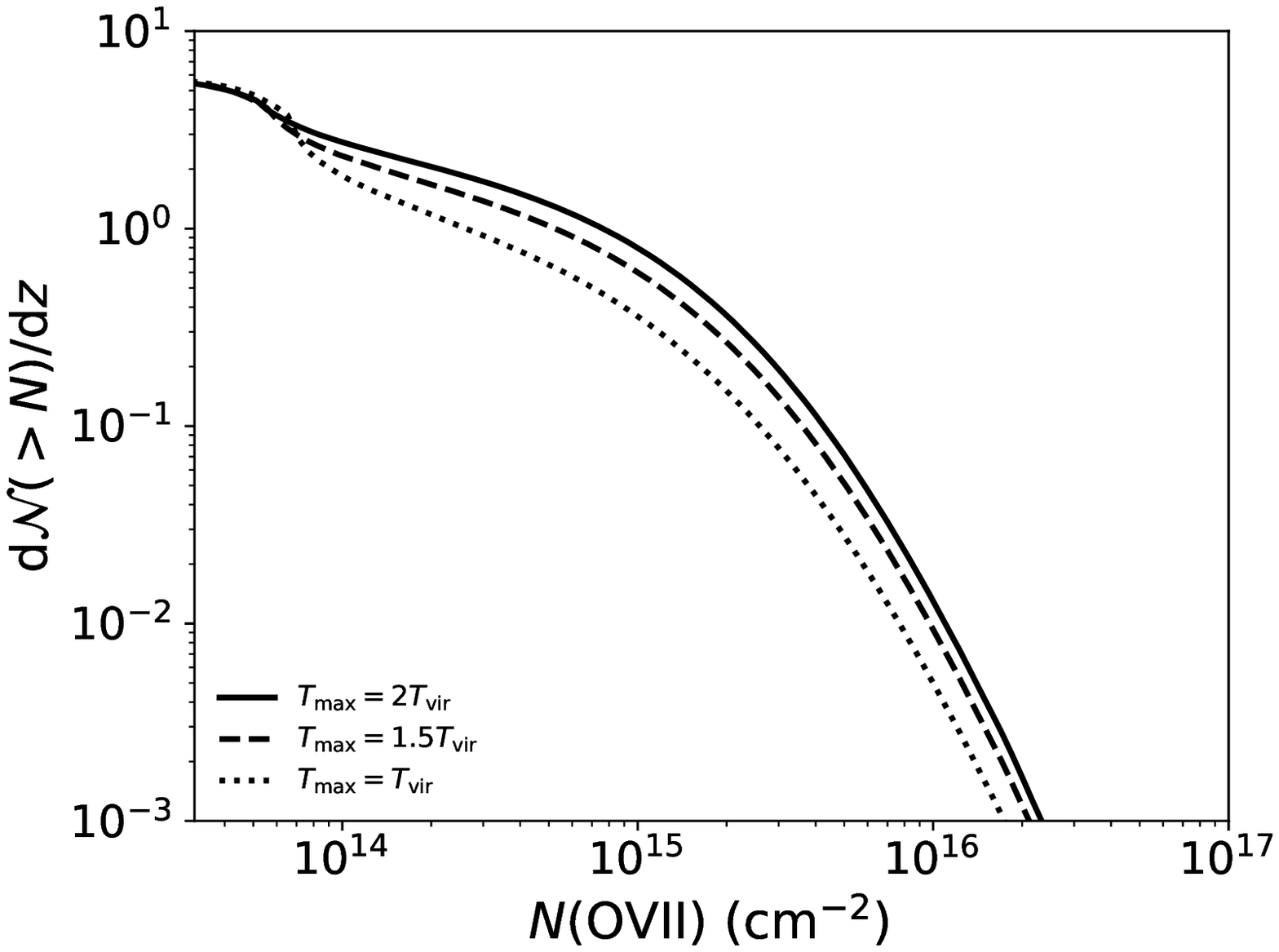}
\includegraphics[width=0.48\textwidth]{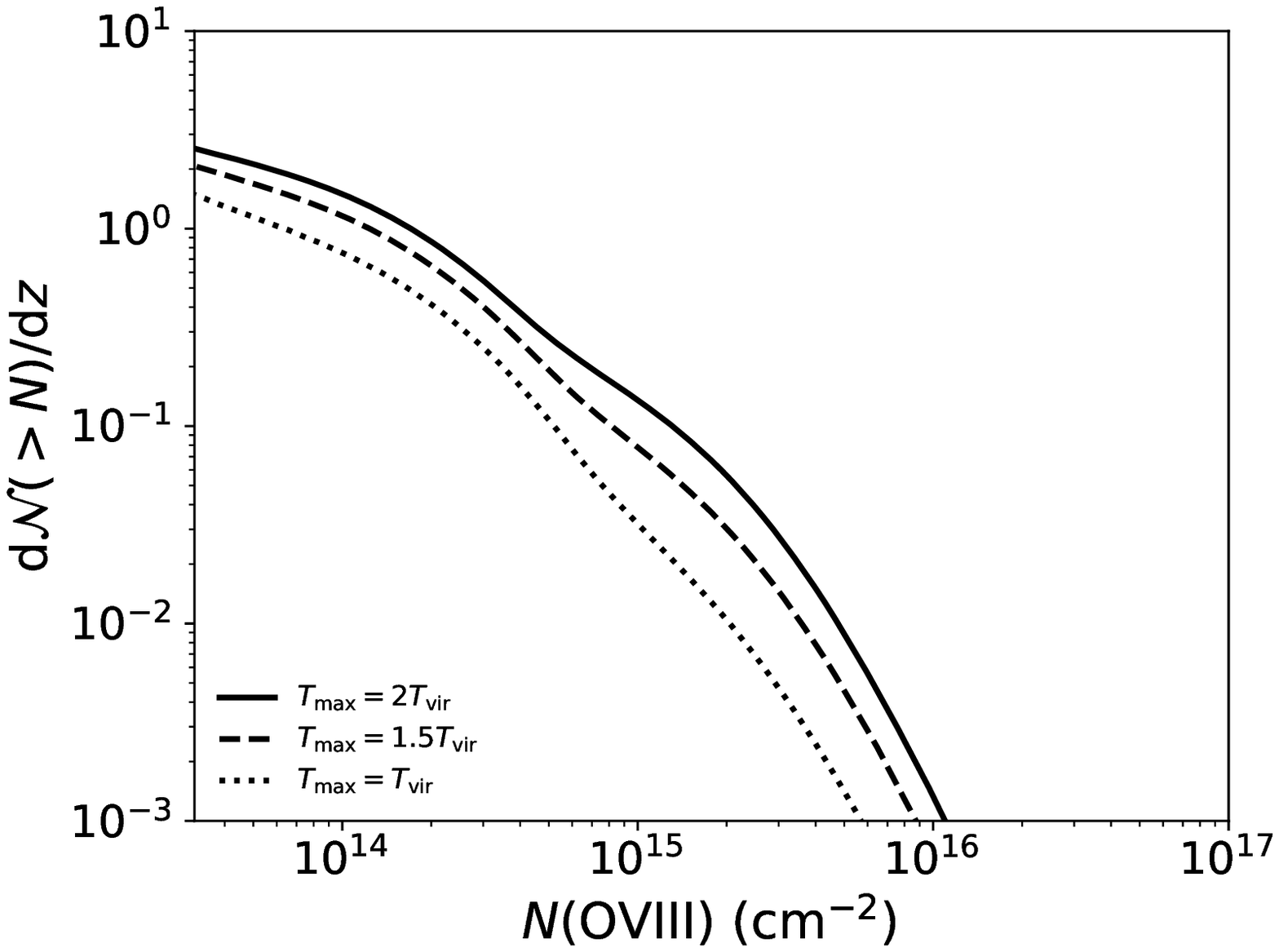}
}
\end{center}
\caption{The effect of varying maximum temperature on {\OVII} ({\it Left panel}) and {\OVIII} ({\it Right panel}). Increasing the maximum temperature could increase the detection rate for {\OVII} and {\OVIII} by a factor of $0.2-0.3\rm~dex$ at high column densities.}
\label{OVII_T}
\end{figure*}

\begin{figure*}
\begin{center}
\subfigure{
\includegraphics[width=0.48\textwidth]{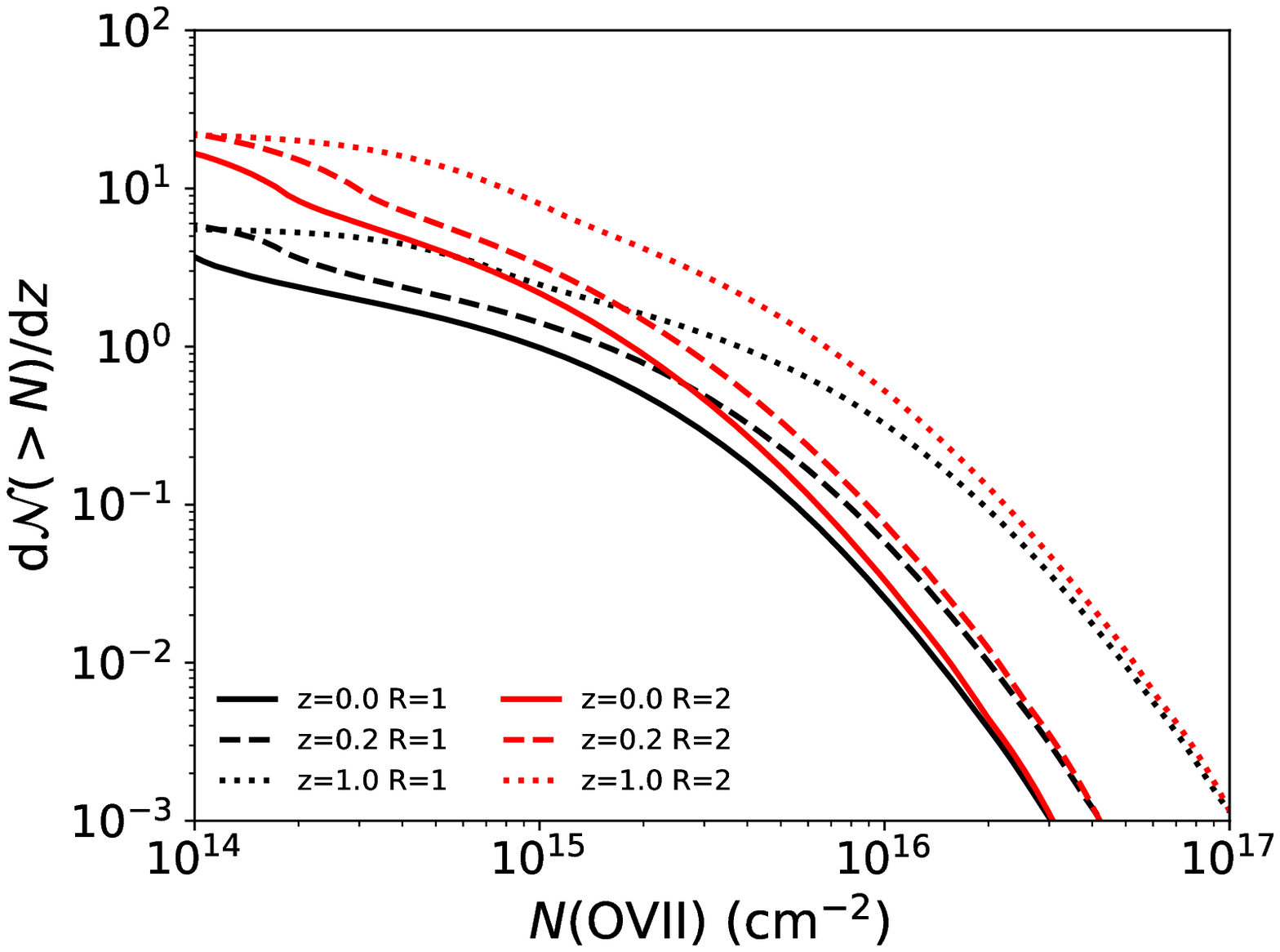}
\includegraphics[width=0.48\textwidth]{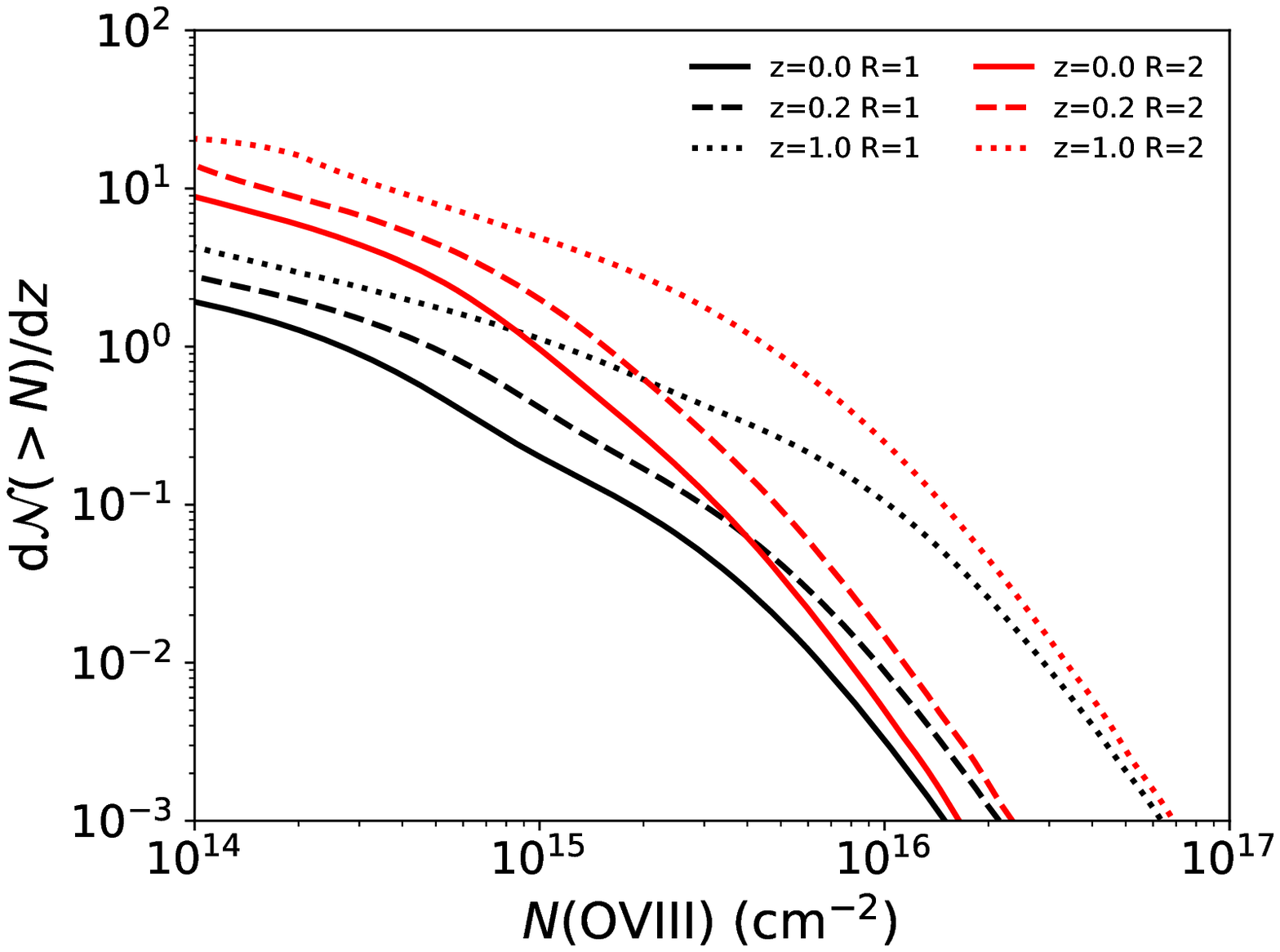}
}
\end{center}
\caption{The redshift evolution of {\OVII} ({\it Left panel}) and {\OVIII} ({\it Right panel}). The predicted column detection rate of {\OVII} and {\OVIII} are about $2-20$ and $0.2-7$ per unit redshift with limiting EW of $1\rm~ m\AA$ at $z=0.0-1.0$.}
\label{OVII_z}
\end{figure*}

The total model detection rate of galactic gaseous halos ($\approx 6$ per unit redshift) is far below the observed detection rate of {\OVI} systems ($\approx 40$  per unit redshift at $\log N = 12.8$; \citealt{Danforth:2016aa}).
Increasing the maximum radius can increase the halo cross section; therefore, we consider models with the maximum radius of twice the virial radius, which is shown in the right panel of Fig. \ref{TZR}.
With a larger maximum radius, the low column density system ($\log N({\rm OVI}) < 14$) is raised significantly by factor $>2$, while the high column density system detection rate is only increased by several percent and is still significantly underestimated. 
At low column densities, the difference cannot be made up completely by the extension of the maximum radius. 
The gaseous halo has a non-zero lower bound for the column density (see Fig. \ref{halo_model}), which means that the differential distribution would decrease to zero at a given column density. 
However, the observed column density distributions do not show such a decrease.
Therefore, we suggest that the cosmic filaments (gas not associated with galaxy halos) might be another origin for low {\OVI} column density systems, which is discussed in Section 4.4.

Finally, we consider the redshift evolution of the {\OVI} column density distribution.
Since the redshift dependence of sSFR  is approximately $\propto (1+z)^3$, the high redshift ($z\gtrsim 1$) leads to a significant higher SFR, which affects a gaseous halo by increasing the density (QB18).
This higher density occurs because the higher rate of star formation requires a higher cooling rate of mass from the halo.
The increase of the gaseous halo density is approximately the root square of the sSFR, therefore a significant increase of the {\OVI} column density is expected for high redshift galaxies (Fig. \ref{OVI_z}).
\red{The $z=1.0$ column density distribution fits the observed distribution in the high column density end phenomenally, although this fitting is non-physical, because the observation samples have smaller redshifts ($z\approx0.2$).
Since the redshift mainly affect the SFR, the $z=1.0$ model is equivalent to a higher SFR model (see discussion in Section \ref{scatter}) or a stronger feedback model (larger $\gamma$ than our assumption from a galactic wind model) at low redshifts.}


\subsection{{\OVII} and {\OVIII} in the Local Universe}
The {\OVII} and {\OVIII} ions show similar predicted column density distributions to {\OVI}, while the break point is moved to a higher column density about $10^{15}\rm~cm^{-2}$.
This change of the break point is mainly because these two ions have higher ionization fractions than {\OVI} and are associated with more massive galaxies, which normally have higher total hydrogen column densities.
Increasing the metallicity will move the distribution toward higher column densities, and increasing the maximum radius raises the detection rate of low-column density systems.

\begin{figure*}
\begin{center}
\subfigure{
\includegraphics[width=0.48\textwidth]{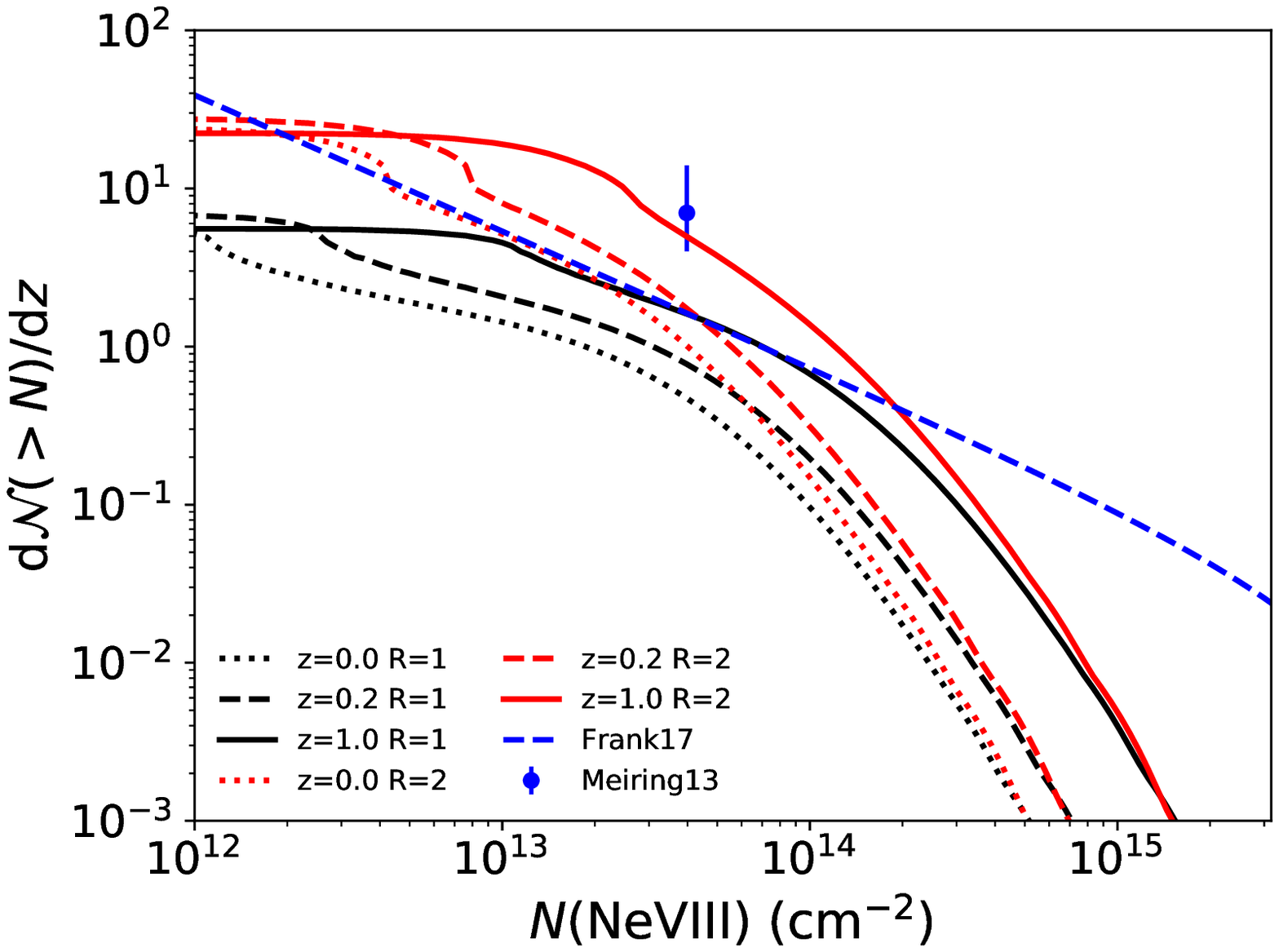}
\includegraphics[width=0.48\textwidth]{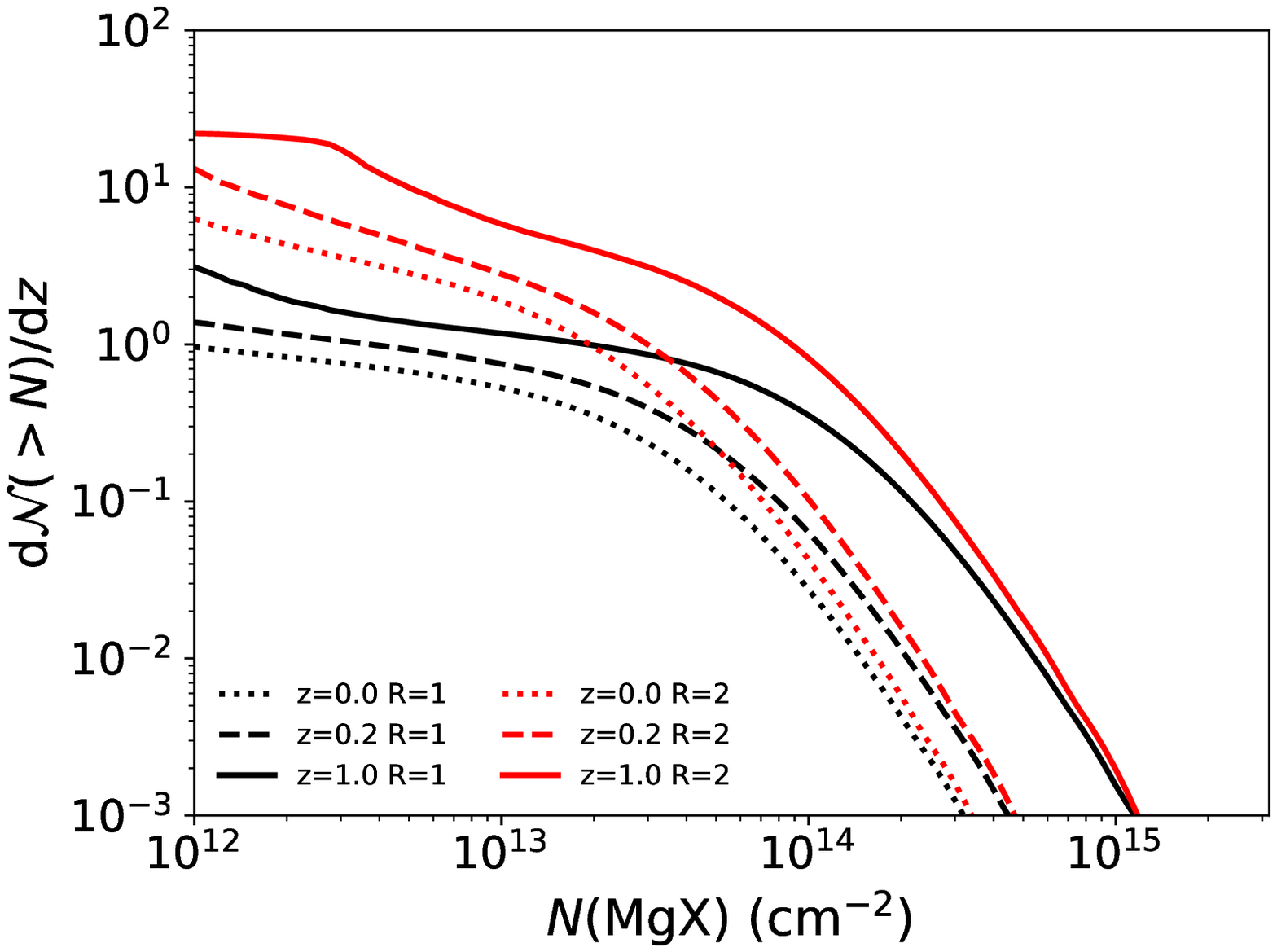}
}
\end{center}
\caption{The redshift evolution of {\NeVIII} ({\it Left panel}) and {\MgX} ({\it Right panel}). The predicted column detection rate of {\NeVIII} and {\MgX} are about $2-5$ and $1-3$ per unit redshift with limiting column density of $\log N = 13.7$ at $z=1.0$. Current observations are not consistent for \ion{Ne}{8} at $\log N = 13.4$, an issue that needs to be resolved in order to constrain the models.}
\label{NeVIII_z}
\end{figure*}

Increasing the maximum temperature increases the detection rates of both {\OVII} and {\OVIII} at high column densities, which is due to the higher ionization potentials of these two ions. 
Once the maximum temperature is increased,  {\OVII} and {\OVIII} occur in lower mass galaxies, which have higher number densities and larger detection rates (Fig. \ref{gh_cf}).
Specifically, the abundance of {\OVII} detections are $\approx 0.3\rm~dex$ higher for high column density systems ($> 10^{15}\rm~cm^{-2}$), and this effect is more significant for {\OVIII} systems ($\approx 0.7\rm~dex$).

Fig. \ref{OVII_z} shows the redshift evolution of the {\OVII} and {\OVIII} column density distribution, which is similar to the {\OVI} column density distribution in Fig. \ref{OVI_z}.
With the maximum radius varying between $1$ or $2~R_{\rm vir}$, our model predicts the detection rate of {\OVII} is $1-10$ in the redshift range $z=0.0-1.0$ with a limiting column density of $10^{15}\rm~cm^{-2}$ (EW $ \approx 3\rm~m\AA$), or $2-20$ with a limiting column density of $3\times 10^{14}\rm~cm^{-2}$ (EW $ \approx 1\rm~m\AA$). 
For {\OVIII}, the predicted detection rate is $0.2-7$ in the redshift range $z=0-1$ with a limiting column density of $10^{15}\rm~cm^{-2}$ (EW $ \approx 1\rm~m\AA$).

\subsection{Intervening {\NeVIII}/{\MgX} near $z=1.0$}
\label{NeVIII_MgX}
The {\NeVIII} and {\MgX} column density distributions have similar characteristics as {\OVII} and {\OVIII}, since {\NeVIII} and {\MgX} have similar ionization potentials to {\OVII} and {\OVIII}, respectively. 
Fig. \ref{NeVIII_z} shows the redshift evolution and the effect of extending the maximum radius for these two ions. 
Our models predict the detection rate of {\NeVIII} and {\MgX} are about $1-2$ and $0.4-1$ systems per unit redshift, respectively, at $z=1$ with a limiting column density of $10^{14}\rm~cm^{-2}$, which requires a continuum $S/N \geq 15$  at a spectral resolution of {\it HST}/COS.

For {\NeVIII}, there are observational studies on the detection rate that can constrain the GGHM models. 
\citet{Meiring:2013aa} shows three {\NeVIII} absorption systems in the sightline of PG 1148+549 around $z=0.7$ with matched {\OVI} absorption features.
Based on this sightline, the derived {\NeVIII} detection rate is ${\rm d}\mathcal{N}/ {\rm d}z = 7^{+7}_{-4}$ with a limiting column density of $10^{13.7}\rm~cm^{-2}$. 
However, the detection rate from \citet{Meiring:2013aa} seems to be inconsistent (at the $2\sigma$ level) with a stacking {\NeVIII} study \citep{Frank:2018aa}. 
The non-detections in the stacked {\NeVIII} spectrum set an upper limit of the {\NeVIII} column density distribution; otherwise, the weaker doublet line should show up in the stacked spectrum.
Meanwhile, individual detections of high {\NeVIII} column density systems set the lower limit of the {\NeVIII} abundance; otherwise, there should not be so many detected systems. 
Combining these two constraints, \citet{Frank:2018aa} gives a preferred single power law distribution of the {\NeVIII} column density, and the detection rate at $z=0.88$ is ${\rm d}\mathcal{N}/ {\rm d}z = 1.4_{-0.8}^{+0.9}$ with $\log N > 13.7$.

As shown in Fig. \ref{NeVIII_z}, these two measurements can constrain the maximum radius of the gaseous halo model.
The extended maximum radius model is consistent with (\citealt{Meiring:2013aa}; slightly lower by a factor of $0.2\rm~ dex$). 
The stacked spectrum study prefers the fiducial model with the maximum radius set to the virial radius \citep{Frank:2018aa}, although the single power law distribution significantly differs from our model at high column densities, which shows a break around $\log N = 13.6$. 
However, for the detectable {\NeVIII} systems ($14.5 > \log N > 13.6$), the fiducial model is consistent with the observations.

\citet{Frank:2018aa} suggested that the {\NeVIII} detections in PG 1148+549 are due to unresolved large cosmic structures, which leads to a high detection rate of {\NeVIII}. 
Therefore, the measurement in \citet{Meiring:2013aa} is an upper limit for the detection rate of {\NeVIII}. 
However, the individual detected {\NeVIII} systems used by \citet{Frank:2018aa} are incomplete (e.g., only the most prominent system is reported in PG 1206+459), and so underestimate the constraint of the lower limit of the {\NeVIII} detection rate. 
Therefore, we suggest that the true value is between the two measurements of \citet{Frank:2018aa} and \citet{Meiring:2013aa}. 
Resolving this discrepancy will require a large and complete survey of {\NeVIII} systems.
Current observations only constrain that the the maximum radius of the hydrostatic gaseous halo is between the virial radius and twice the virial radius.

\section{Discussion}
\subsection{The Effect of the SFR Scatter}
\label{scatter}
The SFR is a crucial parameter to determine the density of the gaseous halo in the GGHM model and the density is approximately proportional to the square root of the SFR (QB18).
In the previous calculation, we use the typical SFR from the star formation main sequence \citep{Morselli:2016aa}, which is the logarithmic mean value at different stellar masses. 
Once the SFR scatter is considered, the arithmetic mean is larger than the typical SFR used in our model, which leads to variations of the final column density distribution.

\begin{figure}
\begin{center}
\includegraphics[width=0.48\textwidth]{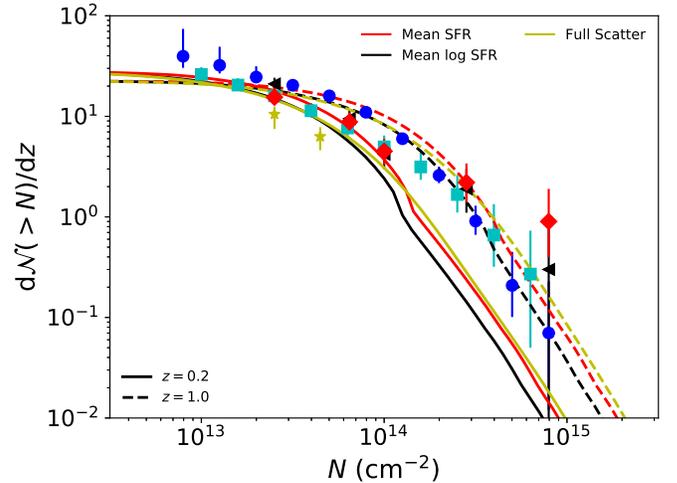}
\end{center}
\caption{The effects of SFR scatter. The observation points are encoded in the same colors as Fig. \ref{TZR}. The black lines are the fiducial models using the logarithm mean of SFR from the star formation main sequence. The red lines use the arithmetic mean instead of the logarithm mean, while the yellow lines are the convolution results (see the text for details).}
\label{OVI_SFR}
\end{figure}

The accurate way to account for the SFR scatter is by integrating over the SFR-$M_\star$ plane instead of fixing the SFR.
However, this integration is extremely computer intensive, so we assumed that the magnitude of the scatter is same at different stellar masses, and follows a log-normal distribution for star-forming galaxies.
Then, we use two ways to estimate the effect of the SFR scatter. 
First, we use the arithmetic mean of SFRs instead of the logarithmic mean.
Because of the log-normal distribution, the arithmetic mean is larger than the logarithmic mean by a constant factor of $10^{\sigma/2}$, where $\sigma$ is the standard deviation of the log-normal distribution. 
For $L^*$ galaxies ($\log M_\star=10.5$), the standard deviation is about $0.4\rm~dex$ \citep{Renzini:2015aa}, which is applied to all galaxies, leading to $1.6$ times larger SFR values.
The results of this modification are shown in Fig. \ref{OVI_SFR}, which focuses on {\OVI}. 
Compared to the logarithmic mean, the new column density distribution moves to higher column densities.
For high column density systems ($\gtrsim 10^{14}\rm~cm^{-2}$), our modification increases the detection rate by a factor about $1.7$, while for low column density systems, the rate hardly changes because they are limited by the geometrical covering factor of galaxy halos.

Another way to estimate the SFR scatter is by using the approximation that the column density is proportional to the square root of the SFR, so $N\propto x^{1/2}$, where $x$ is the ratio between the SFR and the logarithmic mean SFR. 
Then, a convolution is calculated to estimate the final column density distribution,
\begin{equation}
\mathcal{F}_{\rm final}(N) = \int_0^\infty \frac{1}{\sqrt{2\pi \sigma^2}} e^{-\frac{\log x^2}{2\sigma^2}} \times \mathcal{F}(Nx^{-1/2}) {\rm d} x,
\end{equation}
where $\mathcal{F}(N)$ is the column density distribution calculated in Equation 1, and $\mathcal{F}_{\rm final}(N)$ is the final column density distribution with the SFR scatter considered. 
This final column density distribution is shown in Fig. \ref{OVI_SFR}. 
This full scatter estimation shows a relationship that is similar to the first treatment (i.e., the arithmetic mean) within a difference of $0.1~\rm dex$.
At all redshifts, the convolution solution has higher detection rates than the arithmetic mean solution for high column density systems.
This is because the high SFR wing leads to higher column density systems.
The low SFR wing affects the distribution in the opposite direction, which decreases the detection rate in the column density range near the break point.
Meanwhile, the sharp decrease around the break point is smoothed by the convolution of the SFR, which is more consistent with the observations.

When accounting for scatter in the SFR, the detection rate of high column density systems is increased.
Although our model at $z=0.2$ is still below measurements obtained around $z=0.1-0.7$ \citep{Danforth:2016aa}, the model at $z=1.0$ lies above the observation, which indicates that involving the SFR scatter moderates the tension between the observation and our models.

\subsection{Collisional Ionization or Photoionization}
The physical conditions of high ionization state ions are important for understanding the properties of the hot gaseous components in the universe. 
However, high ionization state ions can be produced either through collisional ionization (CI) at high temperatures or photoionization (PI) at low densities, so we must determine the ionization mechanism.

Observationally, the line width ($b$ factor) of different ions help to define the physical conditions \citep{Tripp:2008aa, Savage:2014aa}.
Using the matched {\HI} and {\OVI} line components, the thermal and the non-thermal velocity components can be decomposed by assuming a single temperature model.
Subsequently, the temperature is derived from the thermal velocity and the physical conditions are determined by comparison with the collisional ionization temperature of {\OVI} ($\approx 10^{5.5}\rm~K$).
Employing a threshold of $\approx 10^5 \rm~K$ ($< 10^{5.3}\rm~K$ for PI and $> 10^{4.7}\rm~K$ for CI), about the half of {\OVI} (or less) are collisionally ionized \citep{Savage:2014aa}.

This method is only applied to the ion {\OVI}, since {\HI} and {\OVI} have comparable wavelengths for their resonant lines. 
For higher ionization state UV ions (i.e., {\NeVIII} and {\MgX}), it is very difficult to obtain similar quality spectra for both {\HI} and these two ions, since {\NeVIII} and {\MgX} have much shorter wavelengths, which puts {\HI} in the COS/NUV band with lower resolution and sensitivity.
An alternative way to determine the physical condition is to check whether a single phase PI model could reproduce high ionization state ions. 
Most previous studies show that the observed {\NeVIII} column density cannot be generated in PI models, which require unrealistically low densities and hence extremely large sizes ($> 1\rm~Mpc$; \citealt{Savage:2005aa, Narayanan:2012aa, Meiring:2013aa, Qu:2016aa, Pachat:2017aa}).

Recently, \citet{Hussain:2017aa} argued that PI models can reproduce all {\NeVIII} systems using the UVB prescription  from \citet{Khaire:2015ab}.
Compared to \citet{Haardt:2012aa}, the UVB in \citet{Khaire:2015ab} is $0.5\rm~dex$ higher at and above the {\NeVIII} and {\MgX} ionization edges ($E>10\rm~Ryd$). 
This modification increases the density in the absorbing system for the PI model and reduces the size of the absorption gas to $\lesssim 100\rm~kpc$. 
However, the UVB at the high energy band is mainly from QSOs, which is uncertain observationally. 
\citet{Khaire:2015ab} used the {\it HST}/COS QSO composite spectrum from \citet{Stevans:2014aa}, which has a harder spectral index $\alpha = -1.41\pm 0.15$ ($L_\nu \propto ν^{\alpha}$), compared to $\alpha=-1.57\pm.0.17$ \citep[][from {\it HST}/Faint Object Spectrograph]{Telfer:2002aa} used in \citet{Haardt:2012aa}. 
However, both of these observations only go down to $500\rm~\AA$ ($E=1.8\rm~ Ryd$), which is extrapolated to $25\rm~\AA$ ($E\approx 100\rm~Ryd$) to obtain the UVB at the high energy band. 
Therefore, the actual value of the UVB  is still uncertain in the high energy regime (for {\MgX}, {\NeVIII}, and even {\OVI}).

\begin{figure}
\begin{center}
\includegraphics[width=0.48\textwidth]{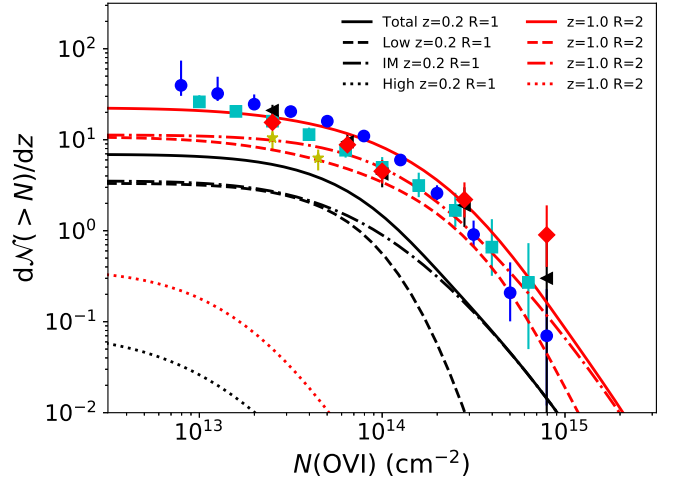}
\end{center}
\caption{The contributions from galaxies with different masses. The observation points are encoded in the same colors as Fig. \ref{TZR}. The three galaxy mass ranges are $\log M_\star = 7.5-8.0$ (low mass -- ``Low"; dashed), $8.0-11.0$ (intermediate mass -- ``IM"; dotted-dashed), and $11.0-12.0$ (high mass -- ``High"; dotted). The intermediate and high mass galaxies are collisionally ionized in GGHM models, while the low mass galaxies are photoionized.}
\label{PIorCI}
\end{figure}

In the GGHM models, PI and CI systems can be distinguished in another way based on statistical properties. 
Phenomenally, {\OVI} absorption systems that originate from PI and CI show different radial dependences as shown in Fig. \ref{halo_model} and Fig. \ref{OVI_imp}. 
This is consistent with observations -- the dwarf galaxy sample shows a flatter radial dependence \citep{Johnson:2017aa}, while $L^*$ galaxies show a radial decrease ($2.9\sigma$), which is defined and dominated by ``broad" features \citep{Werk:2016aa}.
Therefore, we select a threshold for the PI to CI transition at a the stellar mass of $10^8~M_\star$, above which we assume that {\OVI} is predominantly collisionally ionized.

Using this criterion, the contribution of PI and CI are shown for two models: the fiducial model at $z=0.2$ and the extended radius model at $z=1.0$ in Fig. \ref{PIorCI}. 
For these models, there is a significant difference between the upper limit column density for PI and CI {\OVI}, which indicates high column density {\OVI} systems are all collisionally ionized. 
The PI contribution shows a sharp decrease around the break point for both models, which is because the PI {\OVI} leads to in a narrow range of column densities near $\approx 10^{14}\rm~cm^{-2}$ at $z=0.2$ (Fig. \ref{halo_model}).
For the current detection limit of {\OVI} ($\log N({\rm OVI}) \approx 13$), the ratio between CI and PI systems is around $1$ for gaseous halos, so about the same amount of {\OVI} is excited by each process.
However, our model with $R_{\rm max}=2R_{\rm vir}$ still underestimates the detection rate of low column density systems.
Assuming this difference is mainly caused by unvirialized IGM gas that is photoionized (discussed further in Section 4.4), one will obtain a CI/PI ratio about $0.3$ when the CI contribution is fixed to our gaseous halo model.

\begin{figure*}
\begin{center}
\subfigure{
\includegraphics[width=0.48\textwidth]{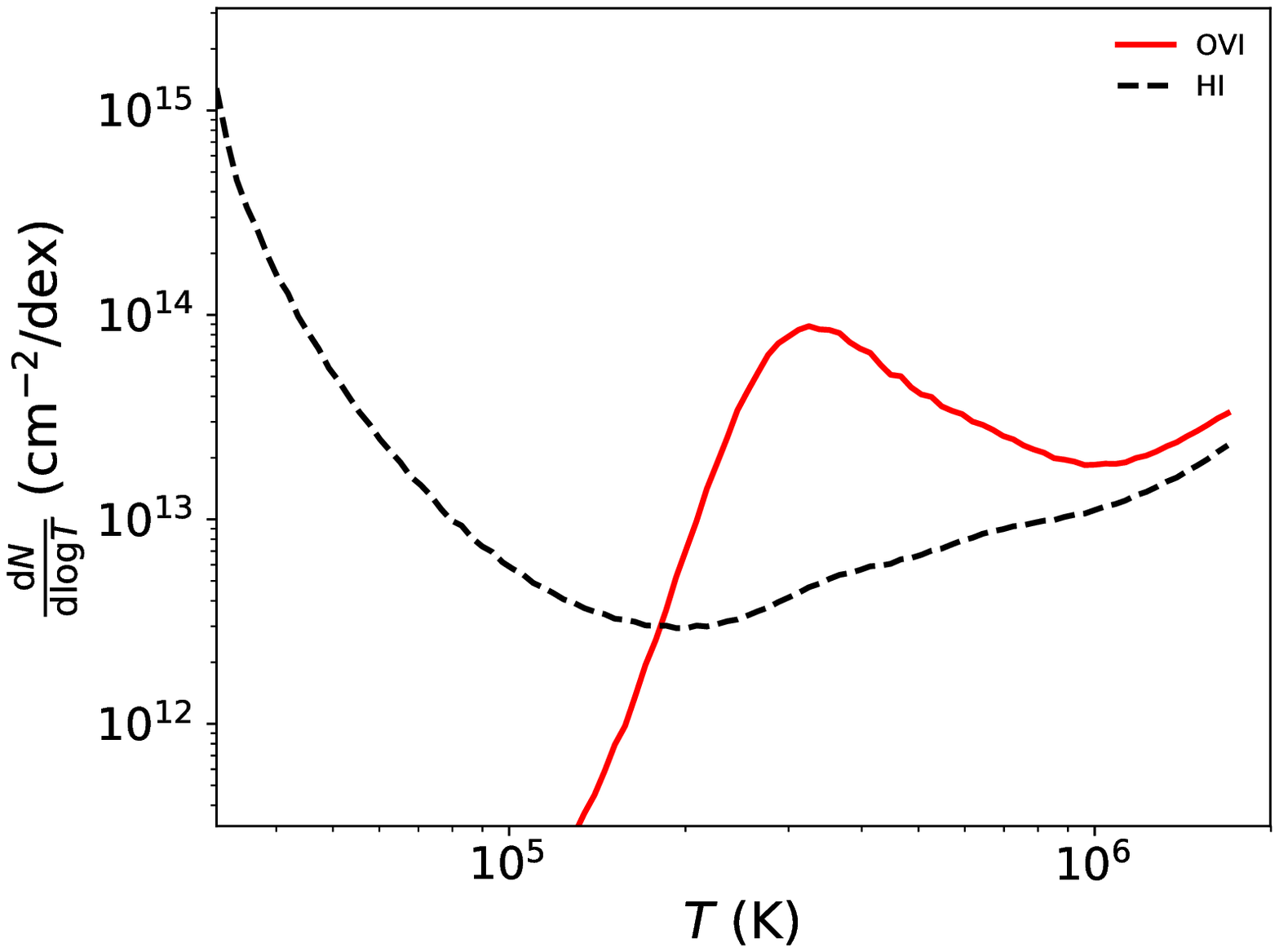}
\includegraphics[width=0.48\textwidth]{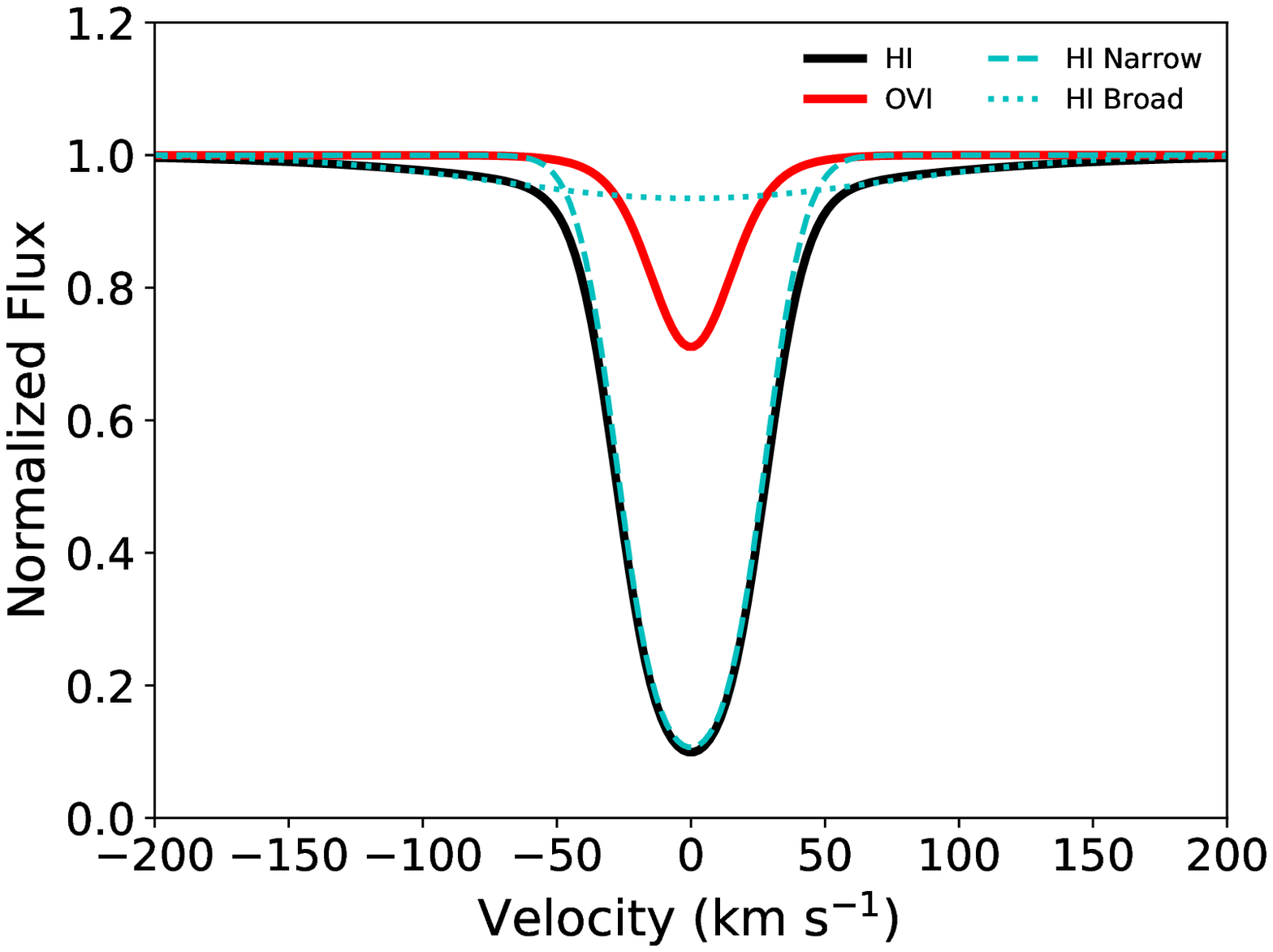}
}
\end{center}
\caption{Example of the multi-phase medium in the gaseous halo, with galaxy parameters $M_\star = 10^{10.5}M_\odot$, ${\rm SFR} = 3 M_\odot~\rm yr^{-1}$, $Z= 0.3Z_\odot$, $R_{\rm max} = 2R_{\rm vir}$, $T_{\rm max} = 2T_{\rm vir}$ at $z=0.2$. The impact parameter of this absorption is $0.3R_{\rm vir}$. {\it Left panel}: The column density dependence on the temperature. {\it Right panel}: The composite line shapes for Ly$\alpha$ and the strong line of {\OVI}. The Ly$\alpha$ line can be decomposed to two components -- broad ($\log T=5.8$) and narrow ($\log T=4.5$).}
\label{line_shape}
\end{figure*}

This CI/PI ratio ($\approx 0.3$) is consistent with observations \citep{Savage:2014aa}; however, our model and the observations have different criteria for distinguishing CI from PI. 
\citet{Savage:2014aa} set a threshold temperature of $\log T=4.7$ for CI, whose {\OVI} ionization fraction is $<10^{-10}$ in collisional ionization equilibrium (CIE). 
If we use $\log T=5.0$ as the threshold, the observed CI/PI ratio is less than the GGHM model prediction of 0.3.
As stated above, the current method to determine the physical condition is using the Doppler $b$ factor, which requires a detectable broad {\HI} features ($b>70 \kms$) to obtain the high temperature ($T\gtrsim10^5\rm~K$). 
However, these broad {\HI} features are difficult to detect, especially when there are also narrow features at the same velocity.

We consider an $L^*$ galaxy with $M_\star = 10^{10.5}M_\odot$, ${\rm SFR} = 3 M_\odot~\rm yr^{-1}$, $Z= 0.3Z_\odot$, $R_{\rm max} = 2R_{\rm vir}$, $T_{\rm max} = 2T_{\rm vir}$, and the impact parameter is $0.3~R_{\rm vir}$ at $z=0.2$.
In the left panel of Fig. \ref{line_shape}, we show the temperature dependence of the column density for {\HI} and {\OVI}.
The temperature distribution is a combination of two factors -- the mass-temperature distribution of the multi-phase medium (QB18) and the ionization fraction.
It is clear that {\OVI} is mainly from collisional ionization, which leads to the peak around $10^{5.5}\rm ~K$.
The {\HI} columns are mainly generated in a lower temperature phase, while there is also a weak, broad {\HI} feature at high temperature ($\approx 10^6\rm~K$).
With these column density distributions, the composite line shapes are also shown in Fig. \ref{line_shape}, where we only consider the thermal broadening.
The Ly$\alpha$ is decomposed into two components: $\log N_1 = 13.86$ with $b_1 = 24.5\kms$ ($\log T=4.5$); and $\log N_2 = 12.96$ with $b_2 = 102.0\kms$ ($\log T=5.8$), while the {\OVI} strong line is well modeled by a single component model with $\log N = 13.54$ and $b=21.4\kms$ ($\log T=5.7$).

In such a system, although the {\OVI} is collisionally ionized (associated with the broad component of {\HI}), it can be mistaken as being photoionized, if the {\OVI} is assigned to the narrow component.
Unfortunately, the broad component of Ly$\alpha$ is difficult to detect since the central depth is only about $5\%$, which is easily buried by the noise, especially in the situation that it is blended with the narrow component.
This situation may occur in observed systems with comparable {\OVI} and {\HI} $b$ factors, which are marked as PI currently \citep{Savage:2014aa}.

\subsection{Comments on the Origins of {OVI}}

As stated in Section 2.1, there are three different origins of high ionization state ions in our analytic models: PI for low-mass galaxies; CI in ambient gases at the virial temperature; and CI in cooling flows in massive galaxies. Different ions have different halo masses corresponding to their ionization potentials, and the {\OVI} occurs in ambient gases of galaxies with $M_\star \approx 10^9-10^{10}~M_\odot$.

Several recent theoretical works also attempt to understand the origins of {\OVI} \citep{Bordoloi:2017ab, McQuinn:2018aa}.
\citet{McQuinn:2018aa} argued that the cooling flow in the gaseous halo is the main source of {\OVI}, and the authors suggested a large cooling rate ($10-100~M_\odot\rm~yr^{-1}$) and a dense gaseous halo ($P/k \approx 100~\rm cm^{-3}~K$) to account for the observed {\OVI} column density ($10^{14.5}\rm~cm^{-3}$) in the COS-Halos sample ($\approx L^*$ galaxies). 
This massive cooling flow requires energetic feedback processes originating from the galaxy disk to disrupt the cooling flow at the low temperatures ($\approx 10^4\rm~K$).
Although, {\OVI} in GGHM models are also in the cooling flow, it is shown that galactic winds cannot be energetic enough to support such a massive cooling flow if we adopt a galactic wind model from the FIRE simulations (\citealt{Hopkins:2014aa, Muratov:2015aa}; QB18). 

\citet{Bordoloi:2017ab} suggested that all observed intervening {\OVI} could be explained in the CIE model, involving an additional parameter -- the cooling flow velocity.
This flow velocity is a non-thermal velocity applied to the gas diffusion or gas flow, and larger flow velocities could lead to higher column densities. 
Therefore, one can find solutions for every single OVI systems in CIE models and the modeled temperatures are all higher than $10^{5.3}\rm~K$.
However, for narrow features, although it can be modeled as a cooling flow, the total line width may be too large  ($v_{\rm cool}^2+ v_{\rm th}^2 > \Delta v^2$; notations from \citealt{Bordoloi:2017ab}). 

For individual {\OVI} systems, \citet{Werk:2016aa} considered their connections with host galaxies.
Specifically, {\OVI} in COS-Halos are divided into three categories -- ``broad" features ($b\gtrsim 40\kms$), ``narrow" features ($b\approx 25\kms$), and ``no-low" features (no corresponding low ionization state ions; \citealt{Werk:2016aa}).
The ``broad" features show a significant decrease  with increasing radius, and typically have high column densities ($\gtrsim 10^{14.5}\cmsq$), while ``no-low" are also broad features ($b>50\kms$), but typically have lower column density ($\lesssim 10^{14.5}\rm \cmsq$).
We suggest that these two types of {\OVI} could correspond to the $M_\star \gtrsim 10^{8.5}~M_\odot$ galaxies and the difference between ``broad" and ``no-low" features are mainly due to the impact parameter.
In our model, with smaller impact parameters, it is more possible to have a cool medium because of a larger cooling rate (higher density).
Therefore, ``no-low" features are all in the outer region (impact parameter $\rho >50\rm~kpc$) and {\OVI} in inner $50\rm~kpc$ regions all have corresponding low ionization state ions \red{(see Fig. 10 in \citealt{Werk:2016aa}).}

The ``narrow" features have small $b$ factors ($\approx 25 ~\kms$), which are also seen in dwarf galaxies \citep{Johnson:2017aa}. 
In the dwarf galaxy sample, all detected {\OVI} absorption lines are narrow ($b\lesssim 30\kms$), except for one system with $b\approx 60\kms$.
These narrow-feature {\OVI} lines in dwarf galaxies are photoionized and associated with low-density gaseous halos ($\approx 10^{-5}\rm~cm^{-3}$) in our model.
However, in the COS-Halos sample,  the``narrow" features also occur in massive galaxies ($>L^*$). 
In addition, \citet{Werk:2016aa} suggested that the UVB-only PIE model is disfavored by the ratio of $N{(\rm NV)}/N{(\rm OVI)}$.
We suggest that these differences may be due to the contamination from dwarf galaxies in the fields, where the identification of such galaxies is limited by the depth of the current galaxy survey.
The failure of the UVB-only PIE model has two potential solutions -- the uncertainty of the UVB (as discussed in Section 4.2) and the additional radiation sources (the escaping flux from the galaxy disk; \citealt{Werk:2016aa}), which are beyond the scope of this paper.

\subsection{\red{The Contribution of Galaxies and the IGM}}
\label{igm}
\red{Our preferred model is $Z=0.5Z_\odot$, $R_{\rm max} = 2 R_{\rm vir}$, and $T_{\rm max} = 2 T_{\rm vir}$ for the galactic gaseous halo based on the modeling of the observed \ion{O}{6} column density distribution.
Also, the SFR is crucial to determine the gaseous halo density and hence the column density, which leads to a strong dependence of the column density distribution on the redshift. 
Using an approximation to the relationship between the SFR and the ion column densities, we introduce the SFR scatter to show that the observed \ion{O}{6} column density lies between our $z=0.2$ and $z=1.0$ models.}

\red{For the simplicity, we use the constant metallicity for all galaxies with different SFR values and stellar masses.
Observationally, \citet{Prochaska:2017aa} studied the metallicity of circumgalactic medium at $z\approx0.2$ using the COS-Halos sample, finding a median metallicity of $-0.51$ dex with scatter $\approx 0.5$ dex.
Assuming the metallicity distribution is a log-normal distribution, the arithmetic mean of the metallicity will be around $0.5-0.6 Z_\odot$, which is consistent with the preferred model.
This constraint on the metallicity is also consistent with the \ion{O}{6} gas in the Illustris-TNG simulation \citep{Nelson:2018aa}.
A direct metallicity measurement of warm \ion{O}{6} ($\log T > 5$) by \citet{Savage:2014aa} also shows a wide range of the metallicity (about $\rm [Z/H] = -2$ to $0$).
Compared to \citet{Prochaska:2017aa}, the median metallicity is lower with $\rm [Z/H] = -0.96$ for six systems.
Due to the uncertainty of the small number statistics, it is still uncertain whether this sample is in conflicts with our assumption and other phase gases.}

\red{The maximum radius is constrained by the \ion{O}{6} column density distribution, and an extended model ($R_{\rm max} = 2R_{\rm vir}$) is preferred.
However, as stated in Section \ref{NeVIII_MgX}, there is a possible disagreement between \ion{Ne}{8} and \ion{O}{6}, where the stacking \ion{Ne}{8} result supports the fiducial model with the virial radius  (but the direct detection of \ion{Ne}{8} systems goes in the opposite sense).
This disagreement may be solved obtaining more \ion{Ne}{8} observations.
Currently, the \ion{O}{6} observations favor the extended radius model for low-mass galaxies, while for massive galaxies, more \ion{Ne}{8} sample is needed.}

\red{The predicted {\OVI} column density distributions have a strong dependence on the redshift due to the evolution of the cosmic SFR.
However, there is no reported redshift evolution of the shape of the \ion{O}{6} column density distribution at $z<1$, while the detection rate of \ion{O}{6} is found to increase with redshift at the limiting column density of $\log N = 13.4$ \citep{Danforth:2016aa}.
The detection rate is modeled as a power law $(1+z)^\gamma$ with a positive index of $\gamma = 1.8\pm 0.8$ for \ion{O}{6}.
As shown in Fig. \ref{OVI_z}, the \ion{O}{6} detection rate is dominated by the galactic gaseous halo contributions above $\log N = 13.4$, and the detection rates also increase with the redshifts in our model.
In our model, the slope of this dependence is not well constrained, but an estimate of this slope is around 1 using the model points at $\log N = 13.4$ in Fig. \ref{OVI_z}.}

\red{With the preferred galactic gaseous halo model, there are two remaining issues -- the differences in the high-column density region ($\gtrsim 10^{14}\rm~cm^{-2}$) and at low-column densities ($\lesssim 10^{13.5}\rm~cm^{-2}$ for {\OVI}). 
The difference at high column densities may be solved by varying the galaxy properties (i.e., the metallicity or the SFR), while the low column density difference indicates the existence of other origins for the high ionization state ions.
We suggest that the intra-group or cluster medium, and/or the cosmic filaments contribute about a comparable number of absorption systems as galaxies at low column densities.
}

\red{The properties of the intra-group or cluster medium are still not well constrained.
\citet{Stocke:2014aa} suggests that the detected broad Ly$\alpha$ features (with \ion{O}{6}) are associated with the smaller galaxy groups rather than the member galaxies.
However, for a single system within a larger galaxy redshift survey, \citet{Stocke:2017aa} finds that a broad \ion{O}{6} is more likely associated with a single galaxy halo (with an impact parameter of $1.1R_{\rm vir}$ and $\Delta v \approx 50 \kms$).
For the Virgo galaxy cluster, \citet{Yoon:2017aa} found that five of six systems with metal lines (up to \ion{C}{4}) have nearby galaxies within $300\rm~kpc$ and $300 \kms$.
They also concluded that the detected Ly$\alpha$ features are IGM around galaxy clusters rather than the intra-cluster medium \citep{Yoon:2012aa, Yoon:2017aa}.}

\begin{figure*}
\begin{center}
\subfigure{
\includegraphics[width=0.48\textwidth]{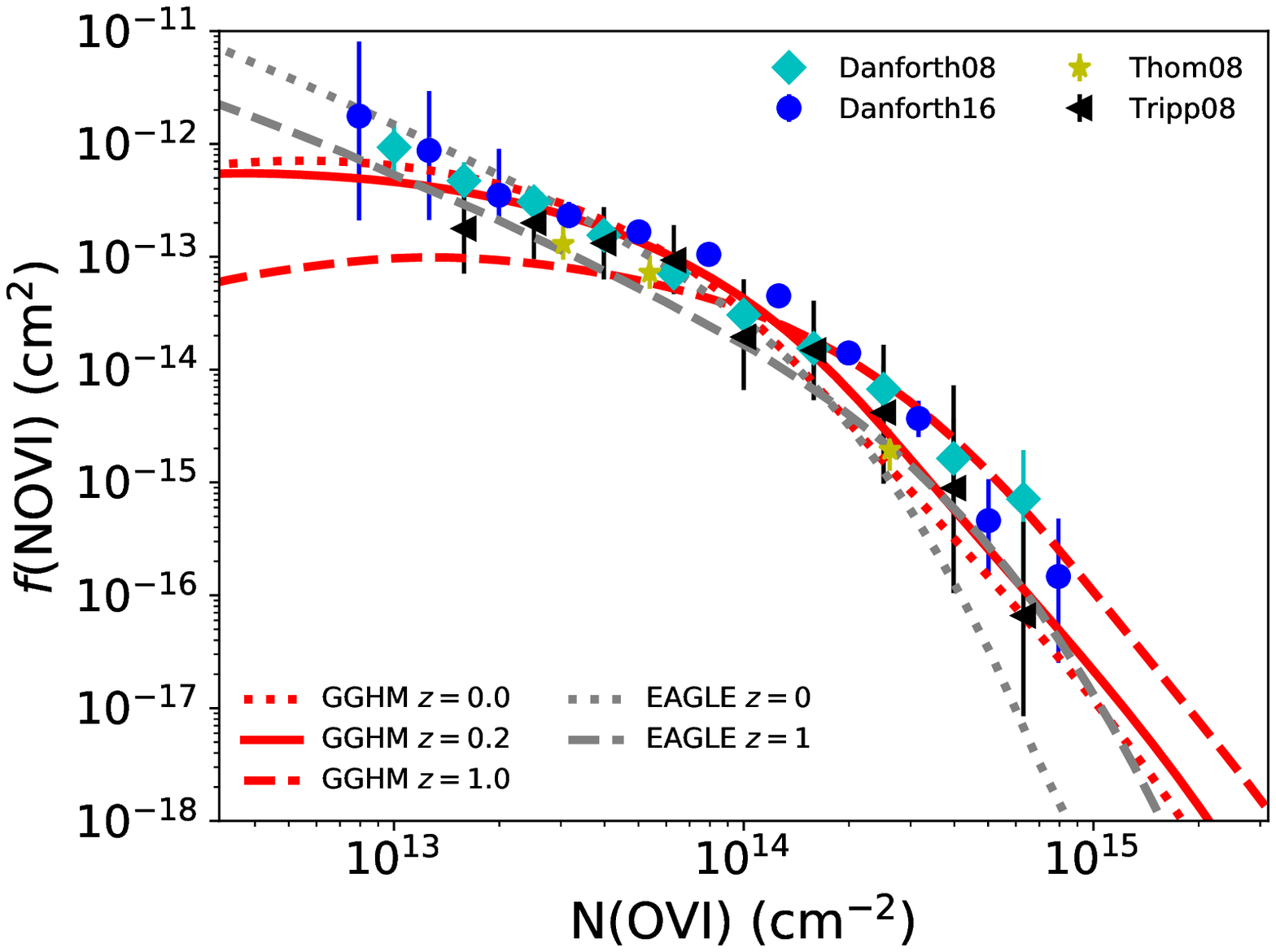}
\includegraphics[width=0.48\textwidth]{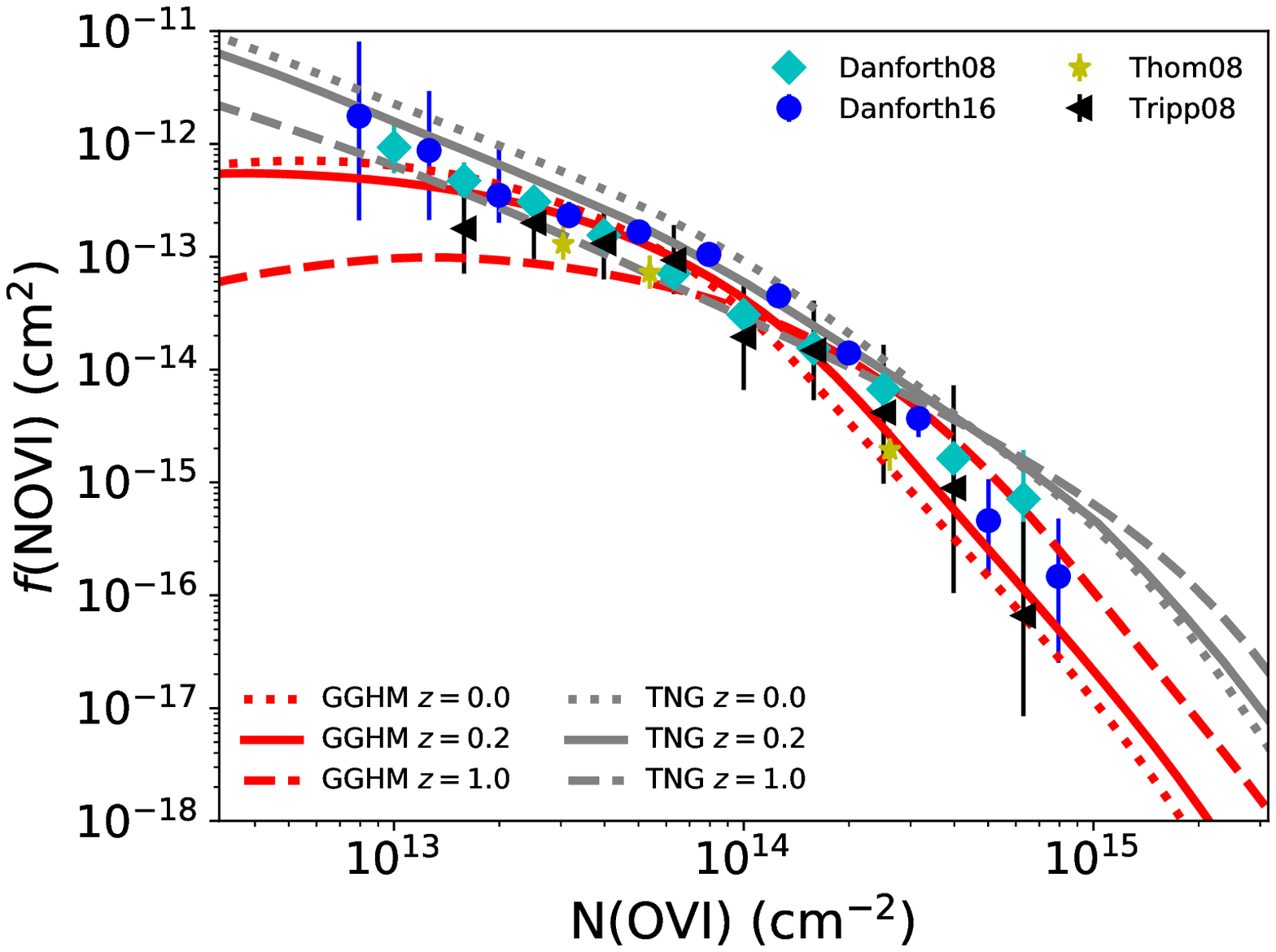}
}
\subfigure{
\includegraphics[width=0.32\textwidth]{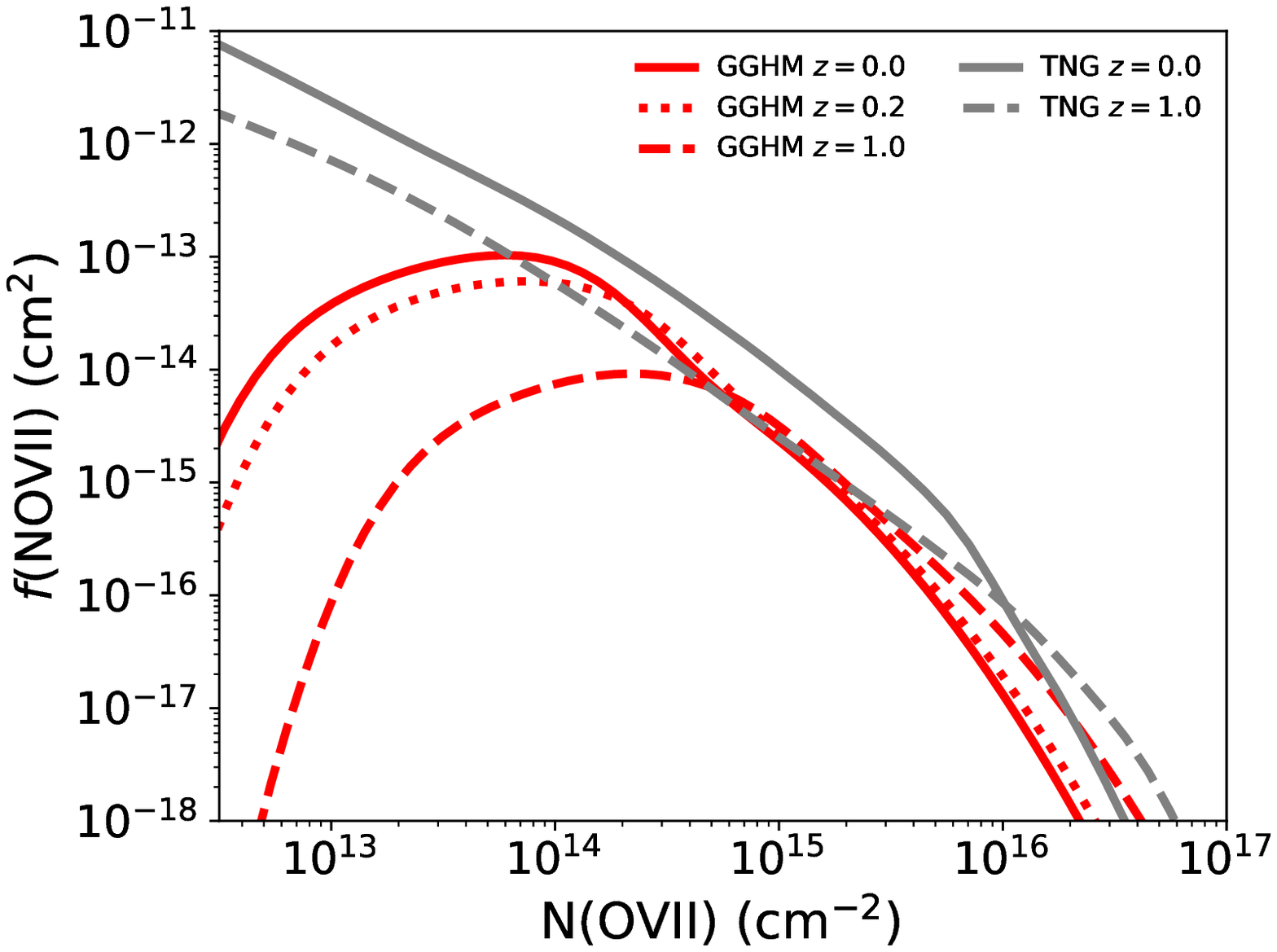}
\includegraphics[width=0.32\textwidth]{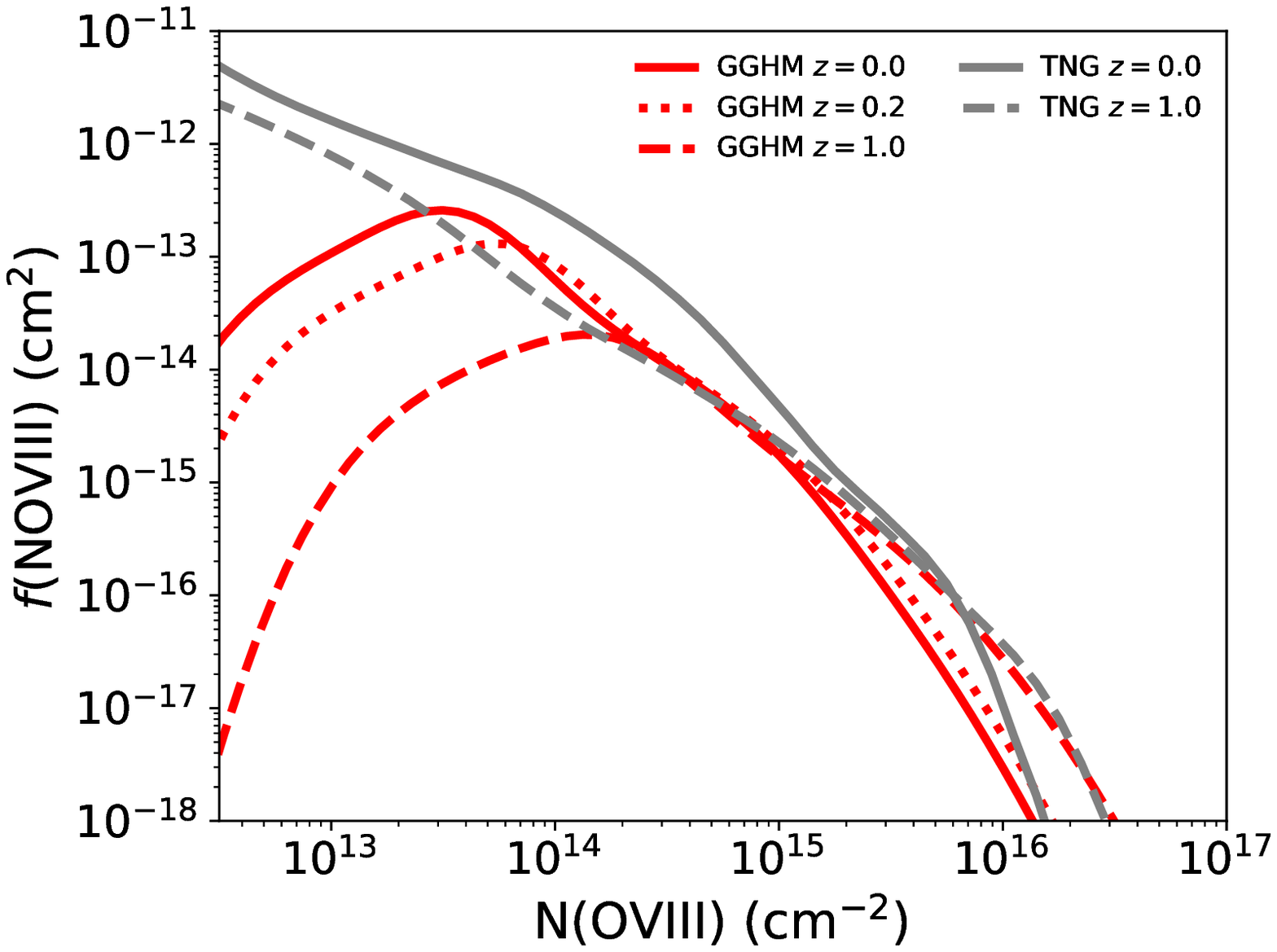}
\includegraphics[width=0.32\textwidth]{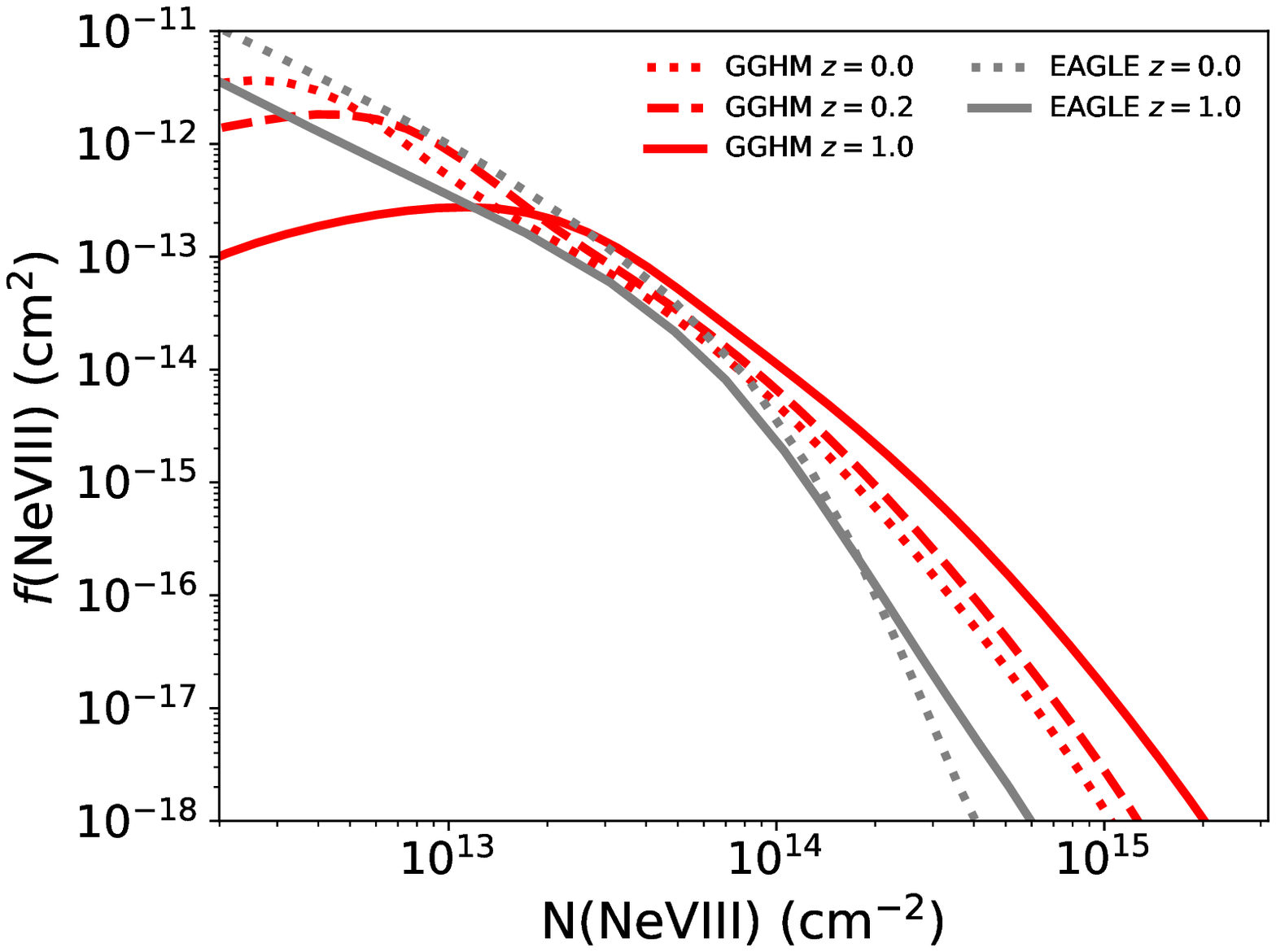}
}
\end{center}
\caption{The comparison of the column density distribution function (Equation 3) between the GGHM models and cosmological simulations, EAGLE and Illustris-TNG. For EAGLE, we compared the {\OVI} and {\NeVIII}, while {\OVI}, {\OVII} and {\OVIII} are available for Illustris-TNG. Three models show comparable column density distributions and similar decrease for high column density systems. The divergence at the low density end indicates the necessity of additional origins for high ionization state ions.}
\label{simu}
\end{figure*}

The properties of cosmic filaments are also rarely constrained in observations due to the difficulty in defining a filament.
\red{\citet{Wakker:2015aa} measured the Ly$\alpha$ for two nearby filaments at $cz \approx 3000\kms$, and found the broad Ly$\alpha$ features ($b> 40 \kms$) are all along the defined filament axes.
However, the corresponding high ionization state ions (i.e., \ion{O}{6}) are out of the wavelength coverage.}
For {\OVI}, one possible sample is from \citet{Savage:2014aa}, where some of the {\OVI} absorption features are beyond twice the virial radius (up to $\approx 5-10~R_{\rm vir}$).
Since the detection limits of the galaxy survey are about $0.01~L^*$, it is still possible that these features are due to smaller galaxies.
Assuming $R_{\rm max}=2R_{\rm vir}$, the detection rate is $\approx 13$ per unit redshift for galaxies with masses $10^{7.5}-10^{8.5}~M_\odot$ ($0.001-0.01~L^*$).
Therefore, we expect that more {\OVI} absorption systems are not associated with galactic halos in this situation, since we require a detection rate of $20$ per unit redshift for cosmic filaments (comparing our models with \citealt{Danforth:2016aa}). 
Based on this assumption, {\OVI} in cosmic filaments has an upper limit of $\approx 10^{14}\rm~cm^{-2}$, obtained from \citet{Savage:2014aa} {\OVI} systems that are $>500\rm~kpc$ from known galaxies.
These values satisfy our requirements of the additional {\OVI} systems beyond the gaseous halo (Fig. \ref{OVI_z}).

\red{The high ionization state ions beyond galactic gaseous halos may have different origins due to their ionization potentials.
It is unlikely that {\OVI} can be the dominant ion in galaxy groups or clusters, since the ionization fraction will be very low ($\lesssim 10^{-4}$) at high temperatures ($\log T > 6.5$).
However, higher ionization state ions (i.e., {\OVII} and {\OVIII}) could be dominant in these large structures. 
Therefore, additional {\OVI} may be from the cosmic web, while additional {\OVII} and {\OVIII} are more likely from galaxy groups or clusters.}

\subsection{Comparison to Galaxy Simulations}
We also compare our models with cosmological simulations to obtain insights into the contributions beyond galactic gaseous halos.
In Fig. \ref{simu}, we show the comparison with two cosmological simulations -- EAGLE \citep{Schaye:2015aa, Rahmati:2016aa} and Illustris-TNG \citep{Pillepich:2018aa, Nelson:2018aa}. 
Although there are several systematic differences, our model and these simulations are similar in that low-column density systems have a larger detection rate at $z=0$, while high-column density systems are more abundant at $z=1$ for all high ionization state ions. 
From the high$-z$ universe to the local universe, the decrease of high column density systems is associated with the decreasing cosmic SFR in our model, which is consistent with the evolution from $z=2$ to $z=0$ in simulations \citep{Rahmati:2016aa, Nelson:2018aa}.

In our model, the increasing detection rate of low-column density {\OVI} features is due to the low SFR at $z=0$. 
However, as shown in Fig. \ref{simu}, the galaxy contribution does not dominate the low column density end for all high ionization state ions; therefore, it is necessary to consider the unvirialized gases (cosmic filaments), which is discussed in Section \ref{igm}. 

There are some systematic differences between our model and simulations, which might be explained by the variation of the SMF.
We employed the observed SMF, while EAGLE and Illustris-TNG both produce their own SMF obtained from the simulations. 
The EAGLE SMF is lower than the observation by a factor of $0.2\rm~dex$ in the range of $M_\star = 10^{10}-10^{11}~M_\odot$ (the major contributor of {\NeVIII}; \citealt{Schaye:2015aa}), which could lead to a lower {\NeVIII} column density distribution than our models at high column densities.
For the Illustris-TNG, the {\OVI} is overestimated at high column densities mainly due to the $0.2\rm~dex$ higher spatial density in their simulations for galaxies with $M_\star = 10^{9}-10^{10}~M_\odot$ \citep{Pillepich:2018aa}. 
Besides the difference in the SMF, different treatments of physical processes also lead to different column density distributions. 
Our model predicts $0.1-0.2\rm~dex$ more baryonic material than EAGLE for galaxies of $M_\star = 10^{10}-10^{11}~M_\odot$ (QB18), which also contributes to the difference in the {\NeVIII} column density distribution. 
However, this kind of comparison is too small to distinguish between current models, so we do not include such a comparison.

\section{Summary}
Following QB18, we calculate the column density distribution of high ionization state ions (i.e., {\OVI}, {\NeVIII}, {\OVII}, {\MgX}, and {\OVIII}) originating from galactic gaseous halos. 
We convolved the GGHM models from QB18 with the redshift-dependent SMF from \citet{Tomczak:2014aa} to obtain the redshift evolution of the column density distributions.
We summarize our major results as follows:
\begin{enumerate}
\item In the GGHM model, there are three processes that lead to high ionization state ions: photoionization and collisional ionization in the virialized halo; and a radiative cooling flow that forms around sufficiently massive galaxies. Collisional ionization becomes dominant for a halo mass whose virial temperature corresponds to the excitation potential of a given ion.

\item \red{Our model reproduces the \citet{Johnson:2015aa} and \citet{Johnson:2017aa} samples at $z\approx 0.2$, while underestimates the COS-Halos sample at the same redshift but with higher masses, which is discussed in QB18.}

\item Observationally, the \ion{O}{6} column density distribution is modeled as a broken power law \citep{Danforth:2016aa}. The power law at high column densities could be reproduced by the galaxies around the transition mass of {\OVI} ($M_\star = 3\times 10^9~M_\odot$; see Fig. \ref{OVI_z} and Fig. \ref{PIorCI}).  \red{Based on \ion{O}{6} observations, a typical gaseous galaxy halo has $T_{\rm max} = 2T_{\rm vir}$, $R_{\rm max} = 2 R_{\rm vir}$ and $Z = 0.5Z_\odot$. The predicted column density distributions are shown in Fig. \ref{simu}.}

\item In our models, the differential column density distributions have turnovers at low column densities  (Fig. \ref{simu}), which is because gaseous halos have the non-zero lower limits of the column densities for all high ionization state ions. However, such turnovers are not observed for {\OVI} and do not occur in simulations. Therefore, additional hosts are required to account for the gaps between our models and observations or simulations.
We suggest that these additional contributions are from the cosmic filaments (IGM) and the intra-group/cluster medium. The non-galactic {\OVI} systems are more likely to be associated with the IGM due to its low ionization potential, although the physical properties of IGM are  still highly uncertain. Additional {\OVII}, {\OVIII}, {\NeVIII}, and {\MgX} could originate in galaxy groups or the outer parts of poor clusters, which contribute to the total column density distributions.

\item The extended radius model ($R_{\rm max} = 2 R_{\rm vir}$; for low-mass galaxies) is favored by the both the absorption-galaxy pair sample and intervening column density distribution studies of {\OVI}. This model may be in tension with the current {\NeVIII} observation studies, although the column density distribution is still uncertain and requires more observations. Combining the {\OVI} and {\NeVIII} column density distributions could give more constraints on the maximum radius of the gaseous halo.

\item The total {\OVI} column density distribution is not sensitive to the temperature of the ambient gas, although the ambient temperature affects the {\OVI} column densities in individual galactic gaseous halos. The detection rate of {\OVII} and {\OVIII} are raised by a factor of $0.3-0.5$ dex at the limiting column density of $10^{15}\cmsq$ (Fig. \ref{OVII_T}), which could be distinguished by future X-ray observations.

\item In GGHM models, the redshift evolution of the column density distribution is dominated by the cosmic $\rm SFR$ at high column densities, which is consistent with cosmological simulations and \ion{O}{6} observations \citep{Danforth:2016aa}.

\end{enumerate}

\acknowledgments
The authors would like to thank Eric Bell, Edmund Hodges-Kluck, and Hui Li for thoughtful discussion and assistance. Z. Q thanks for Dylan Nelson for sharing Illustris-TNG data with us and Sean Johnson for the dwarf galaxy sample. Support for this work is gratefully acknowledged from NASA through ADAP program grants NNX16AF23G and NNX15AM93G. We gratefully acknowledge support from the Department of Astronomy at the University of Michigan.

\bibliographystyle{apj}
\bibliography{MissingBaryon}
\end{document}